\documentclass[acmsmall]{acmart}
\usepackage[ruled]{algorithm2e}
\usepackage{multirow}
\usepackage{tabularx}
\AtBeginDocument{%
  \providecommand\BibTeX{{%
    \normalfont B\kern-0.5em{\scshape i\kern-0.25em b}\kern-0.8em\TeX}}}

\setcopyright{acmlicensed}
\copyrightyear{2018}
\acmYear{2018}
\acmDOI{XXXXXXX.XXXXXXX}

\acmJournal{JACM}
\acmVolume{37}
\acmNumber{4}
\acmArticle{111}
\acmMonth{8}

\begin{document}

\title{Knowledge Graph Pruning for Recommendation}

\author{Fake Lin}
\email{fklin@mail.ustc.edu.cn}
\orcid{0009-0003-1402-2358}
\affiliation{%
  \institution{University of Science and Technology of China}
  \city{Hefei}
  \country{China}
}

\author{Xi Zhu}
\email{xizhu@mail.ustc.edu.cn}
\affiliation{%
  \institution{University of Science and Technology of China}
  \city{Hefei}
  \country{China}}

\author{Ziwei Zhao}
\email{zzw22222@mail.ustc.edu.cn}
\affiliation{%
  \institution{University of Science and Technology of China}
  \city{Hefei}
  \country{China}
}

\author{Deqiang Huang}
\email{deqianghuang@mail.ustc.edu.cn}
\affiliation{%
 \institution{University of Science and Technology of China}
 \city{Hefei}
 \country{China}}

\author{Yu Yu}
\email{yuyu9601@outlook.com}
\affiliation{%
 \institution{Alibaba Group}
 \city{Hangzhou}
 \country{China}}
 
\author{Xueying Li}
\email{xiaoming.lxy@alibaba-inc.com}
\affiliation{%
  \institution{Alibaba Group}
  \city{Hangzhou}
  \country{China}}

\author{Zhi Zheng}
\email{zhengzhi97@mail.ustc.edu.cn}
\affiliation{%
  \institution{University of Science and Technology of China}
  \city{Hefei}
  \country{China}}
\author{Tong Xu}
\authornotemark[1]
\email{tongxu@ustc.edu.cn}
\affiliation{%
  \institution{University of Science and Technology of China}
  \city{Hefei}
  \country{China}}

\author{Enhong Chen}
\email{cheneh@ustc.edu.cn}
\authornotemark[1]
\affiliation{%
  \institution{University of Science and Technology of China}
  \city{Hefei}
  \country{China}}

\renewcommand{\shortauthors}{Fake Lin, et al.}

\begin{abstract}
Recent years have witnessed the prosperity of knowledge graph based recommendation system (KGRS), which enriches the representation of users, items, and entities by structural knowledge with striking improvement. Nevertheless, its unaffordable computational cost still limits researchers from exploring more sophisticated models. We observe that the bottleneck for training efficiency arises from the knowledge graph, which is plagued by the well-known issue of knowledge explosion. Recently, some works have attempted to slim the inflated KG via summarization techniques. However, these summarized nodes may ignore the collaborative signals and deviate from the facts that nodes in knowledge graph represent symbolic abstractions of entities from the real-world. To this end, in this paper, we propose a novel approach called KGTrimmer for knowledge graph pruning tailored for recommendation, to remove the unessential nodes while minimizing performance degradation. Specifically, we design an importance evaluator from a dual-view perspective. For the collective view, we embrace the idea of collective intelligence by extracting community consensus based on abundant collaborative signals, i.e. nodes are considered important if they attract attention of numerous users. For the holistic view, we learn a global mask to identify the valueless nodes from their inherent properties or overall popularity. With the collective and holistic importance scores, we build an end-to-end importance-aware graph neural network, which injects filtered knowledge to enhance the distillation of valuable user-item collaborative signals. Ultimately, we generate a pruned knowledge graph with lightweight, stable, and robust properties to facilitate the following-up recommendation task. Extensive experiments are conducted on three publicly available datasets to prove the effectiveness and generalization ability of KGTrimmer, where it can reduce the number of triplets in KG by up to 90\% without compromising performance. 
\end{abstract}




\maketitle

\section{INTRODUCTION}
Incorporating knowledge graph (KG) as a supplementary resource to support the recommendation system has gained significant popularity. By leveraging KGs, we can obtain a more comprehensive understanding among users, items, and entities, which leads to a better characterization of user preference and item relevance \cite{DBLP:conf/www/WangHWYL0C21,DBLP:conf/kdd/Wang00LC19,DBLP:conf/sigir/0002ZCZG22}. 
KGs can also enhance the transparency and interpretability of recommendation systems, as they provide clear evidence of how we arrive at the recommendation results \cite{DBLP:conf/www/WangHWYL0C21}.

To better leverage the KG data, numerous researchers are dedicated to incorporating users, items, and entities into a shared embedding space based on a collaborative knowledge graph (CKG), which aims to manipulate the structured knowledge and produce explainable recommendation results \cite{DBLP:conf/kdd/Wang00LC19}. 
This line of literature adopts the widespread message aggregation scheme by recursively collecting neighborhood information on the CKG to capture the long-range connectivities \cite{DBLP:journals/aiopen/ZhouCHZYLWLS20,DBLP:conf/kdd/Wang00LC19,DBLP:conf/kdd/YangHXH23,DBLP:conf/www/WangHWYL0C21}. Hence, the valuable knowledge information could be distilled from the KG to boost the collaborative signals over the user-item interactions.

\begin{figure*}[t]
\centering
\includegraphics[scale=0.85]{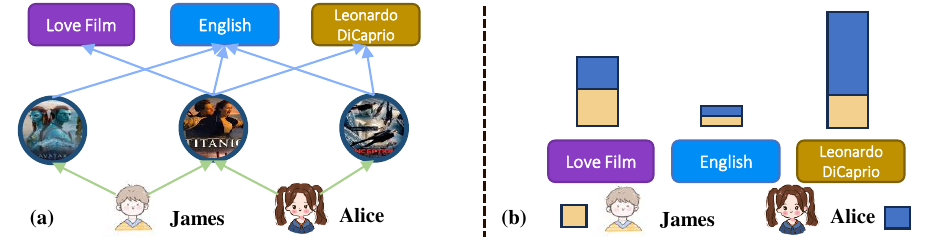} 
\caption{The motivation example from an online a movie platform, where (a) presents a real-world collaborative knowledge graph and (b) indicates the corresponding collective importance of each entity for user modeling.}
\label{fig1}
\end{figure*}

Despite the prevalence, previous literature is based on the steadfast assumption that KGs consistently deliver accurate, reliable, and valuable facts, thereby undoubtedly enhancing recommendation performance \cite{DBLP:conf/www/WangHWYL0C21,DBLP:conf/sigir/0002ZCZG22,DBLP:conf/kdd/YangHXH23}. However, in real-world scenarios, large-scale hand-crafted KGs frequently suffer from evident knowledge redundancy and encounter uncontrolled knowledge quality \cite{DBLP:conf/iclr/XuFJXSD20,DBLP:conf/cikm/WangLFSY21}. This not only imposes additional computational overhead but may even adversely undermine recommendation performance. Inspired by this,  pruning the KGs is indispensable to better serve recommendation systems with improvements on both efficiency and accuracy.

To this end, we investigate the knowledge graph pruning approach tailored for recommendation systems, which is a never-before-studied problem. 
Early efforts has been devoted to KG summarization \cite{DBLP:conf/sigcomm/DoerrB13,DBLP:journals/jiis/SacentiFW22,DBLP:conf/icde/KangLS22,DBLP:conf/iclr/Brody0Y22,DBLP:journals/kbs/GoyalCC20}, employing techniques such as cluster-based inference \cite{DBLP:conf/sigcomm/DoerrB13,DBLP:journals/jiis/SacentiFW22} or GNN-based approaches \cite{DBLP:conf/icde/KangLS22,DBLP:conf/iclr/Brody0Y22,DBLP:journals/kbs/GoyalCC20} to compress the KGs and enhance subsequent utilization efficiency. However, this line of research shares inherent limitations, as the summarized nodes may deviate from the facts that nodes in knowledge graph represent symbolic abstractions of entities from the real-world, thereby impacting the interpretability of the KG. On the contrary, KG pruning aims to obtain a learnable vector to determine the preservation or removal of each triplet, so that we can evaluate their contribution while preserving the original semantics of knowledge. Recently, Graph Sparsification\cite{DBLP:journals/ijdsa/LiZTJFZ22,zhang2024graph,DBLP:conf/uai/YuAYJP22,DBLP:conf/icml/ChenSCZW21} has come into the spotlight, which attempts to remove noisy edges to obtain a robust graph structure. However, studies in this field solely evaluate the importance of individual attributes or triplets from the structure of KG, without any consideration of recommendation task. 
Consequently, the pruning result cannot precisely align with the recommendation system, resulting in suboptimal performance.

In this paper, we target at eliminating the valueless nodes in the KG with respect to recommender system. Formally, we define the nodes that cannot bring any information gain for KG-based recommendation task as valueless nodes. In order to tackle this task, there are two challenges urged to be addressed: 
\begin{itemize}
    \item \textbf{How to evaluate the importance of nodes?} Traditional pruning methods emphasize the structural coherence and connectivity within knowledge graphs, yet such methods are inadequately suited for pruning tasks in KG-based recommendation systems. For example, in conventional pruning tasks, nodes with high degrees are more likely to be preserved , as their removal would greatly compromise the connectivity of the KG \cite{DBLP:conf/ijcai/FaralliFPV18}. In contrast, when it comes to KG-based recommendation tasks, things may be different: collaborative filtering signals are introduced but the majority of literature \cite{DBLP:conf/ijcai/FaralliFPV18,DBLP:conf/sigcomm/DoerrB13,DBLP:conf/cikm/ZhangDDHLX22} fails to integrate it to identify the knowledge that is useless for the accurate modeling of user profiles. Taking Fig \ref{fig1}(a) as an example, as all movies are connected to "English" node, it is incapable of differentiating user preferences and therefore contributes less informational value to the recommender tasks. Hence, we need to formulate an innovative importance criterion that incorporates insights from collaborative filtering signals. 
    \item \textbf{How can we acquire ground truth to inform and direct the pruning process?} To the best of our knowledge,  the mainstream of the works shows limited adaptability to the task. Supervised methods primarily rely on manually introduced erroneous nodes as labels \cite{DBLP:conf/cikm/ZhangDDHLX22}. However, although randomly injected nodes could be perceived as noises or errors, they can hardly be regarded as valueless nodes. Moreover, Unsupervised studies\cite{DBLP:conf/www/BelthZVK20} utilize rule-learning approaches to detect strange or abnormal facts. Nevertheless, applying these method is ill-advised, as the valueless nodes could be retained due to their high occurrence. Fortunately, Some recent studies \cite{DBLP:conf/www/XuDT22} attempt to utilize the gold answers from the downstream tasks to guide the pruning process. In light of this, we are expected to devise a strategy to deliver the signals to pruning tasks.
\end{itemize}

To face the challenges, we propose an adaptive knowledge graph pruning model tailored for recommendation called \textbf{KGTrimmer}, aiming to remove uninformative or valueless knowledge while minimizing performance degradation. 
Initially, we evaluate the contribution of each entity in the KG through an importance evaluator from a dual-view perspective. For the collective view, we embrace the idea of collective intelligence by extracting and aggregating consensus based on collaborative signals, i.e. nodes are considered important if they attract attention of numerous users. As depicted in \ref{fig1} (b), neither "James" nor "Alice" shows interest on "English", then "English" will be tagged as valueless node.    
For the holistic view, we obtain the holistic importance of each entity via a global mask to identify the valueless nodes from their inherent properties or overall popularity. By aggregating the importance scores of the collective and holistic views, we establish an aggregated mask vector that automatically identifies the essential and indispensable entities in the KG. Afterward, we build an end-to-end importance-aware graph neural network, which simultaneously exploits user-item interactions and the item-item relevance with the filtering mechanism on the CKG. 
Upon identifying the important nodes with the iterative process, we ultimately generate a valuable, stable, and robust KG after careful knowledge pruning in the context of the given recommendation scenarios. We make the following main contributions:
\begin{itemize}
\item We are among the first ones to develop knowledge graph pruning approaches specifically tailored for personalized recommendation scenarios, intending to expedite the training process while not compromising predictive accuracy.
\item We propose a novel model called KGTrimmer. Specifically, we automatically select the essential and valuable entities in a dual-view perspective, i.e., collective and holistic views. Then, we design an importance-aware graph neural network with constraint information flow to distill the informative knowledge for user modeling.
\item We conduct extensive experiments on three public benchmarking datasets to validate the superiority of our proposed KGTrimmer. The pruned KG preserves the most essential facts and better serves the recommendation system. 
\end{itemize}




\section{RELATED WORK}

\subsection{Knowledge-Aware Recommendation}
Recent knowledge-aware recommendation literature can roughly fall into three categories: path-based, embedding-based and GNN-based models. 

\noindent
\textbf{Path-based models} \cite{DBLP:conf/recsys/CatherineC16,DBLP:conf/kdd/HuSZY18,DBLP:conf/www/MaZCJWLMR19,DBLP:conf/recsys/Sun00BHX18,DBLP:conf/cikm/WangZWZLXG18,DBLP:conf/aaai/WangWX00C19} focus on the patterns of the long-connectivity among users and entities via items. Using RippleNet \cite{DBLP:conf/cikm/WangZWZLXG18} as an illustrative example, this model initiates with users as seeds as disseminates their information in a ripple effect to gather the preferences of all related nodes. Nonetheless, the intricate nature of the graph structure renders RippleNet prone to information explosion. Other methods, such as meta-path \cite{DBLP:conf/kdd/HuSZY18,DBLP:conf/kdd/JinQFD00ZS20}, are delicately designed to capture the user preference on certain entities, and thus shed some light on the interpretation. However, these schemes are gradually fading out due to the heavy dependency on pattern quality.

\noindent \textbf{Embedding-based models } \cite{DBLP:journals/algorithms/AiACZ18,DBLP:conf/www/0003W0HC19,DBLP:conf/sigir/HuangZDWC18,DBLP:conf/sigir/WangZMLM20,DBLP:conf/kdd/ZhangYLXM16} attempt to incorporate the semantic information in knowledge graph. They equip items with the KGE (knowledge graph embedding) techniques to discover latent relationships among entities. For instance, KTUP \cite{DBLP:conf/www/0003W0HC19} believes there is an implicit translational relationship between users and items, where the differences between them can be identified as user preferences. To this end, KTUP adopts the TransH \cite{DBLP:conf/aaai/WangZFC14} to resolve the N-to-N issue and establishes a multi-task task to align the information on the side of the knowledge graph. Despite achieving impressive performance on recommendation tasks, these models  overlook the higher-order connectivity within the graph.

\noindent \textbf{GNN-based models} \cite{DBLP:conf/www/WangHWYL0C21,DBLP:conf/kdd/WangZZLZLW19,DBLP:conf/sigir/0002ZCZG22,DBLP:journals/corr/abs-2212-10046} target at aggregating messages from neighbors to enrich the ego node embedding. which inherits the propagation mechanism of GNN\cite{DBLP:journals/aiopen/ZhouCHZYLWLS20}. Typically, KGIN \cite{DBLP:conf/www/WangHWYL0C21} proposes an end-to-end fashion to learn the latent user intents and provide more insight on interpretability. KGRec\cite{DBLP:conf/kdd/YangHXH23} proposes a rational-aware masking mechanism to identify the most important information in the KG, which mitigate the influence of the noisy and irrelevant triplets.
Although they have achieved stunning success, all of them suffer from extremely high computational costs due to the cumbersome KG. Along this line, we propose a simple yet effective method to prune the undesirable nodes and maintain comparable performance on various models and datasets.
\subsection{Knowledge Graph Error Detection}
With the growing adoption of LLM-based knowledge graph construction, concerns regarding the quality of these graphs have amplified, leading to an increased emphasis on the development of error detection techniques for graphs. As a result, numerous researchers dedicated significant efforts to studying how to detect erroneous information within knowledge graphs. For example, CKRL\cite{DBLP:conf/aaai/Xie0LL18} integrated the local triplets and global path confidence to detect the possible noises. KGTtm\cite{DBLP:conf/www/JiaXCWE19} borrowed the idea from PageRank and proposed resource allocation mechanism to measure the trustworthiness of triplets. SUKE\cite{DBLP:journals/access/WangNCL21} considered the uncertain information and structure in the knowledge graph to predict the rationality of triplets. REFORM\cite{DBLP:conf/cikm/WangHCWL21} leveraged the meta-learning skills to collect the knowledge from meta-tasks and predict the triplets confidence under few-shot settings. CAGED\cite{DBLP:conf/cikm/ZhangDDHLX22} adopted the multi-view contrastive learning technique which is tailored for the knowledge graph error detection.

Although the aforementioned techniques are effective in recognizing errors and noisy triplets within a KG, they fall short in detecting valueless information. Large KGs are inundated with such valueless triplets which offer no benefit to downstream tasks while consuming valuable computational resources. Additionally, these kinds of information would prevent models from identifying truly valuable triplets. In this paper, we are committed to identifying these valueless nodes to provide a higher-quality graph for downstream tasks.
\subsection{Graph Summarization}
Researchers have devoted themselves to another the to slim the KGs, called \textbf{Graph Summarization}, which compresses the vast KG into a compact one\cite{DBLP:journals/csur/LiuSDK18,DBLP:journals/corr/abs-2302-06114,DBLP:conf/sigcomm/DoerrB13,DBLP:journals/jiis/SacentiFW22,DBLP:conf/www/BelthZVK20,DBLP:conf/icde/KangLS22,DBLP:conf/iclr/Brody0Y22,DBLP:journals/kbs/GoyalCC20}. One brunch of these approaches leverage machine learning or statistical rules to aggregate nodes. Some works \cite{DBLP:conf/sigcomm/DoerrB13,DBLP:journals/jiis/SacentiFW22} adopt clustering techniques to aggregate neighbors into a super node. Despite the prevalence of
these methods, they still are incapable to incorporate the collaborative signals. KGist\cite{DBLP:conf/www/BelthZVK20} tries to capture the meta-path rules according to the Minimum Description Length principle (MDL). By applying these rules, abnormal edges are identified and eliminated. Another stream of works turns to GNN for help\cite{DBLP:conf/icde/KangLS22,DBLP:conf/iclr/Brody0Y22,DBLP:journals/kbs/GoyalCC20}. For instance, PeGaSus \cite{DBLP:conf/icde/KangLS22} optimizes the summarization of large-scale graphs by considering both user preferences and interests, along with the structural properties of the graph. Regardless of remarkable achievements, KG summarization skills still face undesirable limitations. On the one hand, summarized nodes from the aggregated nodes may deviate from the facts that nodes in knowledge graph represent symbolic abstractions of entities from the real-world. On the other hand, they all put emphasis on the role of local neighbors, the signals from the holistic view are greatly neglected. Along this line, we launched an unexplored task knowledge graph pruning for the recommender system and proposed a novel dual-view-based framework to address the two issues above elegantly.
\subsection{Knowledge Graph Pruning}
Knowledge graph pruning techniques have garnered widespread attention due to the knowledge explosion issue in the information overload era\cite{DBLP:conf/sigir/TianYRWWWL21}. Nevertheless, extensive works confine themselves within the paradigm of subgraph extraction \cite{DBLP:conf/icml/TeruDH20, DBLP:conf/iclr/XuFJXSD20,DBLP:conf/www/JoshiU20},  which still retains a significant amount of irrelevant information, leading to wastage of information resources. Recently, Crumb-Trail \cite{DBLP:conf/ijcai/FaralliFPV18} tells the story in a different way, where it refines the taxonomic structure and permanently deletes the irrelevant edges. Notwithstanding, it only considers the structure of KGs, overlooking the role of collaborative signals from recommendation system. 

Furthermore, the demand for efficiency has sparked the development of \textbf{Graph Sparsification}, which simplifies graph structures to improve computational efficiency. Early work focused on graph partitioning and clustering tasks, which rely on the theoretical analysis of graph properties to guide the sparsification process, such as \cite{DBLP:journals/cacm/BatsonSST13,DBLP:conf/swat/AlthoferDDJ90}. With the rise of GNNs, pruning methods have also emerged. UGS \cite{DBLP:conf/icml/ChenSCZW21} seeks lottery tickets\cite{DBLP:conf/iclr/FrankleC19} in both the graph structure and network parameters to prune task-irrelevant edges and model parameters. Subsequently, AdaGLT\cite{zhang2024graph1} proposed a layer-wise pruning approach that integrates the training and pruning processes, significantly enhancing both efficiency and performance. MoE\cite{zhang2024graph} further borrowed the concept of Mixture of Experts\cite{DBLP:journals/neco/JacobsJNH91}, using multiple expert models to evaluate the importance of pruned edges. Although these methods have achieved impressive results in graph pruning, they struggle to effectively incorporate collaborative filtering signals, making it challenging to adapt them to recommendation tasks.
\section{PRELIMINARIES}
In this section, we will begin by introducing the concepts of knowledge graph and KG-based recommendation systems. Then, we will proceed to provide a formal definition of knowledge graph pruning specifically for recommendation systems.

\subsection{KG-based Recommendation System}
A knowledge graph serves as a repository that stores structured information about real-world facts in the form of a heterogeneous graph, encompassing item attributes, taxonomy, and external commonsense knowledge.

\begin{definition}
    \textbf{Knowledge Graph (KG)}. Given an entity set $\mathcal{V}$ and a relation set $\mathcal{R}$, a knowledge graph can be denoted as $\mathcal{G}=\{(h, r, t)\ \mid h, t \in \mathcal{V}, r \in \mathcal{R} \}$. In particular, each triplet $(h, r, t)$ in the KG represents a connection between the head entity $h$ and the tail entity $t$, which is linked by a specific relation $r$. It is noteworthy that $\mathcal{R}$ is comprised of relations from both canonical and inverse directions.
\end{definition}

In recommendation scenarios, KGs have been proven to significantly enhance the intricate relevance among items and extract deep-seated user preferences, thereby improving recommendation performance. Following traditional works, suppose there exists a user set $\mathcal{U}$ ($|\mathcal{U}| = M$) and an item set $\mathcal{I}$ ($|\mathcal{I}| = N$), the historical interactions between users and items (e.g., click or purchase) can be naturally transformed into a \textbf{user-item bipartite graph}. Specifically, the binary interaction matrix is denoted as $Y \in \{0,1\}^{M \times N}$, where each entry $y_{ui} = 1$ indicates that the given user $u \in \mathcal{U}$ has interacted with item $i \in \mathcal{I}$, otherwise $y_{ui} = 0$. Furthermore, we establish an item-entity collection for alignment, denoted as $\mathcal{A}=\{ (i,e) \mid i \in \mathcal{I}, e \in \mathcal{V}\}$, which allows us to align the item $i$ with the entity $e$. Hence, we seamlessly integrate the KG with the user-item bipartite graph and subsequently develop a unified \textbf{collaborative knowledge graph (CKG)} for the downstream tasks.

\begin{definition}
    \textbf{KG-based Recommendation System (KGRS)}. Given the user set $\mathcal{U}$, the item set $\mathcal{I}$, and their observed interactions $Y$, with the complementary knowledge from $\mathcal{G}$, the target of the KGRS task is to learn a function $\mathcal{F}$ capable of ranking the set of items $\mathcal{I}$ that a user $u \in \mathcal{U}$ has the potential to interact with, i.e. Top-K recommendation task. 
\end{definition}

\subsection{Knowledge Graph Pruning for Recommendation}
Here we present the definition of knowledge graph pruning for recommendation task (KGPR) which targets at reducing redundant and uninformative edges in knowledge graph $\mathcal{G}$ for efficient and effective prediction in KGRS. Specifically, the objective of KGPR task is to learn a binary mask vector $\boldsymbol{s} \in \{0, 1\}^{|\mathcal{G}|}$ for all triplets, where the $k$-th edge $(h_k, r_k, t_k)$ will be preserved if ${s}_k = 1$, otherwise removed. To this end, we employ the binary vector $\boldsymbol{s}$ on the original KG to obtain the pruned KG, denoted as $\Tilde{\mathcal{G}} = \mathcal{G} \odot \boldsymbol{s}$. To be specific, the formal definition of KGPR task can be described as: 
\begin{definition}
\textbf{Knowledge Graph Pruning for Recommendation Systems (KGPR).} Given a user set $\mathcal{U}$, an item set $\mathcal{I}$, an interaction matrix $Y$ and a knowledge graph $\mathcal{G}$, we aim to learn a binary function $\mathcal{H}$, s.t.  $\boldsymbol{s} = \mathcal{H}(\mathcal{U}, \mathcal{I}, Y, \mathcal{G})$ and $\boldsymbol{s} \in \{0, 1\}^{|\mathcal{G}|}$, where $\boldsymbol{s}$ is a binary mask vector. Subsequently, we can acquire the pruned KG, denoted as $\Tilde{\mathcal{G}} = \mathcal{G} \odot \boldsymbol{s}$. According to the pruned KG, the objective of KGPR is to maximize the predictive accuracy of the unobserved user-item interactions in \textbf{KGRS}.
\end{definition}

\begin{figure*}[t]
\centering
\includegraphics[scale=0.55]{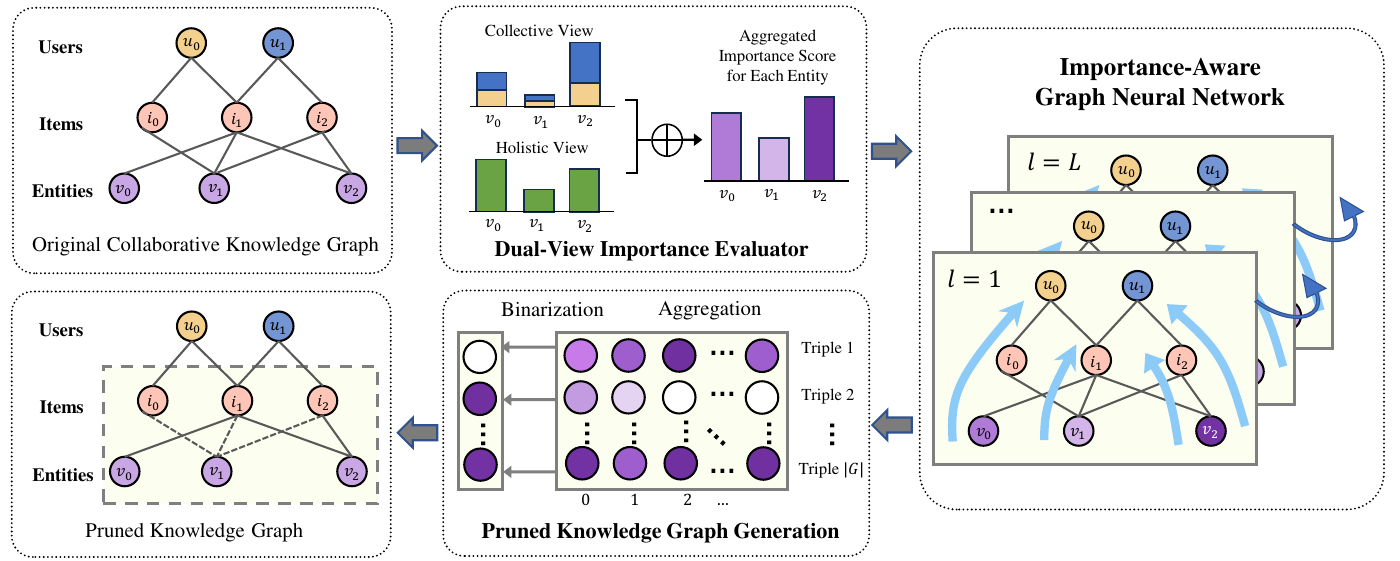} 
\caption{The overall framework of the proposed KGTrimmer framework, which consists three components: (a)Dual-View Importance Evaluator, (b) Importance-Aware Graph Neural Network, and (c) Pruned Knowledge Graph Generation.} 
\label{fig2}
\end{figure*}

\section{METHODOLOGY}
In this section, we begin by introducing the formal formulation of the KG pruning task to gain a comprehensive understanding of overall framework. Following this, we introduce the KGTrimmer model to achieve knowledge graph pruning for recommendation in detail.
\subsection{Overall Framework}
In traditional knowledge graph pruning tasks, researchers often assign a score to each node, which represents its importance. Along this line, we still lay stress on obtaining the final score of each node. In addition, we also aim to maintain the performance of recommendation system after pruning low-informative nodes, i.e. optimize the recommendation system with the pruned knowledge graph. Therefore, there is an urgent need to analyse the task to provide a clearer formula to optimize these two objectives simultaneously. In summary, we formulate the KGPR task as an optimization problem, which is to maximize the probability gap between the positive item $i$ and negative item $j$ for user $u$ given a knowledge graph $\mathcal{G}$, with respect to the estimated mask scores $\hat{\boldsymbol{s}}$:
\begin{align}
\label{eq_overall}
    P_{\theta}(i |u, \mathcal{G})= P_{\theta}(i,|u, \mathcal{G}, \hat{\boldsymbol{s}}) P_{\theta}( \hat{\boldsymbol{s}}|u, \mathcal{G})
\end{align}
where $\theta$ is the parameters of the proposed model, $\hat{\boldsymbol{s}}$ represents the estimated mask scores for all node and $\mathcal{G}$ is the collaborative knowledge graph. Here, $ P_{\theta}(i |u, \mathcal{G})$ means the probability of user $u$ interacting with the item $i$ \cite{DBLP:journals/corr/abs-1205-2618}. Concretely, The latter term $P_{\theta}(\hat{\boldsymbol{s}}|u, \mathcal{G})$ is to estimate the importance of nodes in knowledge graph $\mathcal{G}$, which can be recognized as importance evaluator. The former term $P_{\theta}(i |u,\mathcal{G}, \hat{\boldsymbol{s}})$ is the probability that user $u$ favors the positive item $i$  given the importance score $\hat{\boldsymbol{s}}$ and knowledge graph $\mathcal{G}$, which can be regarded as importance-aware predictor. 

It is worth noting that our ultimate goal is to gather the determined score $\hat{\boldsymbol{s}}$ to locate the valueless nodes, making it rational to prune the knowledge graph in the subsequent process. Along this line, this formulation breaks the conventional knowledge-aware recommendation task into two parts and seamlessly borrows the signal from collaborative information to guide the learning process of scores. Furthermore, the desired score set $\boldsymbol{s} \in \{0, 1\}^{|\mathcal{G}|}$ is the binary vector, whereas our estimated score set $\hat{\boldsymbol{s}}$ is defined as continuous vector for streamlining the gradient calculation in the training process. To bridge this gap, we design a Pruned Knowledge Graph Generation module to binarize the estimated score which serves as the justification for pruning the knowledge graph. In a nutshell, we elaborate on the details of our framework, which is composed of the following modules, as shown in Figure \ref{fig2}:
\begin{itemize}
    \item \textbf{Dual-view Importance Evaluator.} We assess the importance of the candidate entities from a dual-view perspective: collective and holistic view. On one hand, the collective view explicitly utilizes the user attention towards entities to evaluate the importance of entities. On the other hand, the holistic view focuses on the role of the entity within the overall recommendation task. Both contribute to the identification of valueless nodes.
    \item \textbf{Importance-Aware Graph Neural Network.} We leverage the ultimate importance scores to control the information flow for general neural network propagation. In this way, 
    this refined approach allows for the transmission of signals from the recommendation system to the dual-view evaluator, 
    where the module can prioritize crucial information and discard the redundant one. 
    \item \textbf{Pruned Knowledge Graph Generation.} We finally aggregate the candidate masks during the training process and binarize the estimated mask vector to obtain a stable and robust pruned knowledge graph.
\end{itemize}
\subsection{Initialization}
In line with prior work on KG-aware recommendation \cite{DBLP:conf/www/WangHWYL0C21,DBLP:conf/sigir/0002ZCZG22,DBLP:conf/kdd/Wang00LC19}, we assign embedding vector $\boldsymbol{e}^0_u$,  $\boldsymbol{e}^0_v$ and $\boldsymbol{e}_r$ to each user $u \in \mathcal{U}$, entity $v \in \mathcal{V}$ and relation $r \in \mathcal{R}$, respectively. As mentioned above, we omit the item embedding since it is already included in the entity set. Thus, we can construct three matrices:
\begin{align}
    \boldsymbol{E}_u = \{\boldsymbol{e}^0_{u_1}, \boldsymbol{e}^0_{u_2}\,\ldots, \boldsymbol{e}^0_{u_{|\mathcal{U}|}}  \}, \quad
    \boldsymbol{E}_v = \{\boldsymbol{e}^0_{v_1}, \boldsymbol{e}^0_{v_2}\,\ldots, \boldsymbol{e}^0_{v_{|\mathcal{V}|}}  \}, \quad
    \boldsymbol{E}_r = \{\boldsymbol{e}_{r_1}, \boldsymbol{e}_{r_2}\,\ldots, \boldsymbol{e}_{r_{|\mathcal{R}|}}  \}
\end{align}
Here, $|\mathcal{U}|, |\mathcal{V}|, |\mathcal{R}|$ represents the number of the user set $\mathcal{U}$, entity set $\mathcal{V}$ and relation set $\mathcal{R}$, respectively. It is worth noting that, these matrices are initialized via Xavier initializer \cite{DBLP:journals/jmlr/GlorotB10} which guarantees the orthogonality of the initial vectors in the matrix. Taking the user matrix as an example, we have following observations at the initialization step:
\begin{align}
    (\boldsymbol{e}^0_{u_j})^T \boldsymbol{e^0_{u_k}} = 0, \quad j\neq k
\end{align}
where $u_j\in \mathcal{U}$ and $u_k \in \mathcal{U}$ represent any arbitrary users. Subsequently, we employ one-hot vectors $\boldsymbol{ID}$ for each user, entity and relation, and retrieve their corresponding embeddings:
\begin{align}
    \boldsymbol{e}^0_{u}  = \boldsymbol{E}_u^T \cdot \boldsymbol{ID}_u, \quad  \boldsymbol{e}^0_{v}  = \boldsymbol{E}_v^T \cdot \boldsymbol{ID}_v, \quad  \boldsymbol{e}_{r}  = \boldsymbol{E}_r^T \cdot \boldsymbol{ID}_r 
\end{align}
\subsection{Dual-view Importance Evaluator}
In this subsection, we attempt figure out the solution to accurately estimating the score $\hat{\boldsymbol{s}}$, i.e. $P_{\theta}(\hat{\boldsymbol{s}}|\mathcal{G})$.
Along this line, we will introduce how we construct the dual-view KGTrimmer to filter out the effective knowledge that is really beneficial to the collaborative filtering. To be specific, we derive the importance score $\hat{\boldsymbol{s}}$ for each entity from two views, i.e. collective view and holistic view.
For the collective view, we try to consider the user-item interactions to compute the collective importance scores, which show the specific interacted users' preference on the entities. For the holistic view, a learnable mask vector is applied which expects to identify the valueless entities from inherent properties or overall popularity, which is known as the holistic importance scores. After that, we balance the influence of the collective and holistic views and aggregate their importance scores to obtain the aggregated importance scores on all KG entities.


\subsubsection{Collective View} 
 
Users express personalized preferences over items while the items are linked with the entities. Along this line, we realize that the essence of user interest lies in the entities within the KG. Motivated by the concept of collective intelligence, we take advantage of these abundant user-item interactions and consider them as the results of a social discourse. Coupled with the KG, we collect and aggregate their diverse opinions to extract the consensus and determine the importance of each entity for shaping their user profiles in a collective view. 
To this end, we intuitively construct a real-value based user-entity matrix to better represent the correlation between the users and entities. After the construction, each entity receives the attention from all related users to generate a score, a.k.a. collective score. We delve into the details as follows.

\textbf{User-Entity Matrix Construction.} 
 We first gather data from collaborative signals and knowledge graph to assemble the user-entity matrix. For one thing, it is impractical for all entities being aware of the preference of all users, which would result in unbearably high costs. For another, inspired by Monte Carlo algorithm, we do not need to traverse all users related to the entity $v$ to obtain the $\hat{s}_v$. Instead, by adopting a sampling strategy, we can approximate the overall rating of $v$ by examining the scores from the sampled users. Along this line, we follow the prior studies \cite{DBLP:conf/kdd/Wang00LC19} to build a collaborative knowledge graph by combining the user-item bipartite graph and the knowledge graph. For simplicity, we denote the matrix as $Q \in \{0, 1\}^{|\mathcal{U}| \times |\mathcal{V}|}$, where $|\mathcal{U}|$  and $|\mathcal{V}|$ is the number of users and entities respectively. Then we define:
\begin{align}
    Q_{uv}= \begin{cases}1, & \text { there is a path from user } u \text { to entity } v \\ 0, & \text { otherwise }\end{cases}
\end{align}
Next, we employ a more effective Monte Carlo strategy to sample the path from user $u$ to entity $v$. That is to say, if an entity $v$ maintains more than $k$ users, i.e. $|Q_v| > k$, we implement a random sampling strategy to set $Q_{uv}=1$ for the $k$ selected users, while setting $Q_{uv}=0$ for the rest. Here $|Q_v|$ represents the number of paths that end with entity $v$. 

Based on the Monte Carlo algorithm, the mean derived from numerous random samples closely approximates the true distribution's mean. Consequently, by sampling $k$ users, we can estimate the overall user attitude towards this entity, with a larger $k$ resulting in a narrower gap between the estimated and true means. Nevertheless, an increase in k also elevates the computational burden. Thus, a reasonable choice of $k$ is made to optimize the trade-off between precision and computational efficiency.

\textbf{Collective Importance Scores.} 
In this part, we target at revealing the collective importance of the entities in the KG through user-entity matrix. As mentioned above, for each entity $v$, we first refer to the $Q$ matrix and denote the user neighbors of the entity $v$ as $\mathcal{N}_v^Q = \{u | Q_{uv} = 1, u\in \mathcal{U}\}$. Afterward, we collect the opinions of interacted users to compute the collective importance score for the specific entity. Besides, we also enrich the representation of entities by incorporating the information of relation. The formulation can be described as follows:


\begin{equation}
    \begin{aligned}
    \label{equation_1}
        s_v^c &= \frac{1}{|\mathcal{N}_v^Q|} \sum_{u \in \mathcal{N}_v^Q } sim( \boldsymbol{e}_u^0, \boldsymbol{e}_v^\prime), \\
         \boldsymbol{e}_v^\prime &=  \frac{1}{|\mathcal{N}_v^{\mathcal{R}}|}\sum_{r \in \mathcal{N}_v^{\mathcal{R}}} \boldsymbol{e}_r \odot \boldsymbol{e}_v^0, 
    \end{aligned}
\end{equation}
where $\mathcal{N}_v^{\mathcal{R}} = \{r | (h, r, v) \in G\}$, $sim$ is the similarity function, $\odot$ denotes the element-wise product, $\boldsymbol{e}_u^0$, $\boldsymbol{e}_v^0, \boldsymbol{e}_r$ are the embeddings of user $u$, entity $v$ and relation $r$, respectively, and $\boldsymbol{e}_v^\prime$ is the entity embedding which absorbs the information of relations, and $s_v^c$ is the collective importance score for each entity $v$. Formally, we define the similarity function as :
\begin{equation}
    \begin{aligned}
        sim(\boldsymbol{e}_u^0, \boldsymbol{e}_v^\prime) &= max(0,\frac{(\boldsymbol{e}_u^0)^T \boldsymbol{e}_v^\prime}{\Vert \boldsymbol{e}_u^0 \Vert \Vert \boldsymbol{e}_v^\prime \Vert} ),
    \end{aligned}
\end{equation}
We use the modified cosine similarity function for the following considerations: (1) This method naturally limits the output to a range between 0 and 1 and can be considered as a form of dot product followed by normalization. (2) This similarity calculation method ensures that the initial values of the collective score $s^c_v$ are around zero, owing to the fact that the initialization of each matrix is initialized independently. This guarantees that our scores all start from the same point, unlike other normalization methods, such as min-max normalization, which can introduce biases in the scores at the initial stage. Such biases could significantly impact the results of subsequent training. (3) Entities with negative similarity are deemed noisy or erroneous nodes, therefore will be assigned the score of zero and prioritized for removal. Therefore, the collective score is endowed with the ability to jointly identify both noisy and valueless nodes.
Consequently, we obtain the collective importance scores as:
\begin{align}
    \boldsymbol{s}^c = (s_1^c, \dots, s_{|\mathcal{V}|}^c),
\end{align}
which demonstrates interacted users' attention on KG entities based on collaborative signals.

\subsubsection{Holistic View}
The task of KG-based recommendation involves various aspects, such as purchase motivation, product popularity, degree of nodes, etc.
Thus the KG-based recommender system inherently has the potential to identify the holistic importance of entities. To explicitly model this, we assign a learnable parameter to each entity, dubbed as holistic importance score. 

\textbf{Holistic Importance Scores.} Similar to \cite{DBLP:conf/icml/ChenSCZW21}, we define a global mask to represent the underlying value of entities within the KG where the holistic importance scores can be formulated as:
\begin{equation}
    \begin{aligned}
        \boldsymbol{s}^h = (s_1^h, \dots, s_{|\mathcal{V}|}^h),
    \end{aligned}
\end{equation}
To ensure comparability with the collective score, we still limit the range of the holistic importance score between 0 and 1. We have also considered other approaches, such as obtaining specific scores by passing the  entity embeddings through a Multi-Layer Perceptron (MLP). Undoubtedly, MLP possesses a more powerful capability for representation, and thus is commonly used for feature extraction and fusion. However, we still decided to forgo the use of MLP for the following reasons:(1) Effectiveness. In the context of overall framework, the embeddings evolve continuously during the training phase, making it challenging for an MLP to capture their characteristic information effectively. (2) Efficiency. Compared to directly assigning scores to each entity, MLP contains more parameters and requires additional computational resources to support both training and inference.

By leveraging the advantages of the recommender system, the holistic scores possess the capacity to represent the significance of the entities. In other words, entities with low importance scores are ineffective at capturing valuable information for the recommendation task, whereas informative entities with high importance scores are the key drivers in enhancing the performance of the recommendation system.

\subsubsection{Aggregated Importance Scores}
To achieve the knowledge pruning for recommendation, we are desired to perform aggregation operations on these two scores. 
Formally, the aggregated importance scores for entities can be represented as:
\begin{align}
    \hat{\boldsymbol{s}} = \gamma \boldsymbol{s}^c + (1-\gamma) \boldsymbol{s}^h  
\end{align}
where $\gamma$ is a trade-off hyper-parameter. In this way, $\boldsymbol{\hat{s}}$ can be applied for all entities so as to discover the unnecessary or uninformative entities that are expected to be pruned from the KG. Subsequently, we design a transformation function $\mathcal{T}$ to to measure each triplet $(h, r, t)$ according to the $\boldsymbol{\hat{s}}$:
\begin{align}
    \label{score_agg}
    \boldsymbol{\hat{s}}(h,r,t) = \mathcal{T}(\boldsymbol{\hat{s}}_h, \boldsymbol{\hat{s}}_t).
\end{align} 
\noindent where the specific form will be discussed in the Experimental Section.

\subsection{Importance-Aware Graph Neural Network}
In this part, we aim to establish the connection between the recommendation system and the evaluator module, i.e. importance-aware predictor $P_{\theta}(i|u, \mathcal{G}, \hat{\boldsymbol{s}})$. In detail, we inject the final importance scores into the message propagation process of formal KGRS models and make predictions on the unobserved user-item interactions. Previous Graph-based approaches have shown that the neighborhood-based aggregation scheme is a promising method to capture the user preference with the help of KG, such as  KGIN \cite{DBLP:conf/www/WangHWYL0C21} and KGRec \cite{DBLP:conf/kdd/YangHXH23}. Specifically, after broadcasting the importance scores of entities to the corresponding triplets, we emphasize that the key point turns to incorporating the aggregated importance scores into the message propagation process through a dedicated graph neural network and learning the refined representations of users, items, and entities on the CKG. To be specific, we illustrate knowledge graph propagation and user-item graph propagation in detail.


\subsubsection{Knowledge Graph Propagation}
We move on to learning the representations of items and entities with the aggregated importance scores $\boldsymbol{\hat{s}}$. Since some valueless or noisy entities hurt the accuracy or efficiency of KGRS, the importance scores initially provide guidance for knowledge filtering. Formally, we iteratively receive and update the embeddings with constrained information flow in terms of the importance scores as follows:

\begin{align}
\label{kg update}
    \boldsymbol{e}_v ^{(l)} = KGE(\boldsymbol{e}_v^{(l-1)}, \boldsymbol{e}_r, \boldsymbol{\hat{s}}).
\end{align}
Here, we denote $KGE$ as any KG embedding method for entity embedding update, $\boldsymbol{e}_v^{(l-1)}$ and $\boldsymbol{e}_v^{(l)}$ are entity embeddings of previous layer $l-1$ and current layer $l$, respectively. In detail, we adopt the following approach for knowledge graph propagation:
\begin{align}
\label{kg propagation}
    \boldsymbol{e}_v ^{(l)}=  \frac{1}{|\mathcal{N}_v^G|}\sum_{(r, t)\in \mathcal{N}_v^G }  \boldsymbol{\hat{s}}(v, r, t) (\boldsymbol{e}_t^{(l-1)} \odot \boldsymbol{e}_r).
\end{align}
Here $\mathcal{N}_v^G = \{(r, t) | (v, r, t ) \in G\}$ includes the tail entities with their corresponding relation set for a given entity $v$.


\subsubsection{User-Item Graph Propagation}
With the integration of KG, we can delve more profoundly into users' preferences for entities in the KG, which provides more convincing evidence for the recommendation system. Hence, we attempt to apply the filtered knowledge information to enhance the distillation of valuable user-item collaborative signals, leading to a more accurate understanding of user preferences. With the refined item embeddings by knowledge graph propagation, the user embeddings can be calculated as:
\begin{align}
    \boldsymbol{e}_u^{(l)} =CF (\boldsymbol{e}_u^{(l-1)}, \boldsymbol{e}_i^{(l-1)}).
\end{align}
Here $CF$ can be any collaborative filtering method to update user embeddings. $\boldsymbol{e}_u^{(l-1)}$ and $\boldsymbol{e}_u^{(l)}$ are the user embeddings of previous layer $l-1$ and current layer $l$, and $\boldsymbol{e}_i^{(l-1)}$ is the item embedding of previous layer $l-1$. Specifically, we simply adopt the following method to update user embeddings:
\begin{align}
\label{cf}
    \boldsymbol{e}_u^{(l)} = \frac{1}{|\mathcal{N}_u|}\sum_{i \in \mathcal{N}_u}\boldsymbol{e}_i^{(l-1)},
\end{align}
where $\mathcal{N}_u = \{i | Y_{ui} = 1\}$ are the item neighbors of user $u$.

The key distinctions between KG propagation and collaborative graph propagation parts can be summarized as follows: (1) A knowledge graph contains a variety of relationships, necessitating consideration of the effects these relations impart during propagation. In contrast, the collaborative graph is characterized by a singular type of relationship: interaction. This eliminates the necessity for explicit modeling of this interaction to avoid an increase in model complexity. (2) To filter out irrelevant information, message propagation within the KG is weighted by a importance score $\hat{\boldsymbol{s}}(v,r,t)$, which we derived from dual-view evaluator. However, for the CF side, we consider each interaction inherently valuable to the recommendation system and therefore do not perform any filtration. Thus each edge is inherently weighted with a score of 1 to demonstrate the significance of the interaction.


\subsubsection{Prediction}
After stacking $L$ layers, the user embedding $\boldsymbol{e}_u$ and item embedding $\boldsymbol{e}_i$ can be aggregated as:
\begin{equation}
    \begin{aligned}
        \boldsymbol{e}_u &= \boldsymbol{e}_u^{(0)} + \boldsymbol{e}_u^{(1)} + \dots + \boldsymbol{e}_u^{(L)} \\
        \boldsymbol{e}_i &= \boldsymbol{e}_i^{(0)} + \boldsymbol{e}_i^{(1)} + \dots + \boldsymbol{e}_i^{(L)}.
    \end{aligned}
\end{equation}

\noindent where $L$ denotes the number of layers of message propagation. Then we adopt inner product to compute predictive scores and utilize BPR loss to train the recommendation model:
\begin{equation}
\begin{aligned}
    \mathcal{L} = \sum_{u \in \mathcal{U}} \sum_{i \in \mathcal{N}_u} \sum_{j \notin \mathcal{N}_u}^{N_{neg}} \log \left(\sigma\left( \boldsymbol{e}_u^\top \boldsymbol{e}_i - \boldsymbol{e}_u^\top \boldsymbol{e}_j\right)\right),
\end{aligned}
\end{equation}
where $N_{neg}$ is the number of negative samples for each user-item interaction in the training set.

\subsection{Pruned Knowledge Graph Generation}
\label{sec:generation}
In this section, a final pruned knowledge graph $\Tilde{G}$ will be generated for the downstream tasks based on $\boldsymbol{\hat{s}}(h, r,t)$. To achieve this goal, we still be urgent to tackle two challenges: 1) How to aggregate scores from different training stages. In the early stages of training, the model focus more on generic knowledge to fast adapt to the specific task. While in the later stages of training, the model will lay stress on refining details to better fit the training set. As a result, the importance score $\hat{\boldsymbol{s}}$ will evolve as training stages change. Thus, it is worth exploring to find an appropriate to aggregate these scores for the final pruning. 2) How to binarize the final estimated mask vector. Given that our ultimate goal is to produce a pruned knowledge graph, we need to obtain a binarized mask vector to decide which triplets should be retained and which should be discarded. Therefore, to bridge the gap between the obtained scores and the desired mask binary vector, a binarization technique is supposed to be employed.

\textbf{Masked Score Aggregation. }
In this part, our objective is to identify an effective aggregation approach to pinpoint the least valuable information. In fact, during the training process, the candidate mask $\boldsymbol{\hat{s}}^{k}$ can be generated in each epoch. Without loss of generality, the objective mask vector $\boldsymbol{\Tilde{s}}$ can be presented as:
\begin{equation}
\begin{aligned}
    \boldsymbol{\Tilde{s}} = AGG(\boldsymbol{\hat{s}}^{0}, \boldsymbol{\hat{s}}^{1} \dots, \boldsymbol{\hat{s}}^{K}),
\end{aligned}
\end{equation}
where $K$ is number of training epochs, $AGG$ is the function to aggregate all learned mask vectors of each epoch. 

\textbf{Masked Score Binarization. }
As previously discussed, it's essential to transform the scores into a binary format to generate the ultimate pruned graph. The most straightforward strategy involves employing a threshold-based filtering technique, where setting a threshold value $\tau$, allows for the retention of triplets with scores above $\tau$ and the elimination of those below, that is: 
\begin{equation}
    \begin{aligned}
        s_i = \begin{cases}
            1, & \text{if } \Tilde{s}_i \geq \tau \text{ ,} \\0, &\text{otherwise}  
        \end{cases}
    \end{aligned}
\end{equation}
where $i\in\{1,2,\ldots,|\mathcal{G}|\}$ represents the index of triplets. Alternatively, we also implement a percentile-based selection method to retain a certain amount of triplets. In this way, this would involve sorting all scores of triplets and only keeping the desired top $X$ percent of the triplets. Selecting an appropriate threshold facilitates the exclusion of irrelevant nodes by the system. 

Indeed, both methods have their strengths and weaknesses. As for the threshold filtering approach, its advantage lies in removing nodes deemed irrelevant, simultaneously simplifying the dataset while retaining higher-scoring triplets. This method is well-suited for practical application scenarios. For the percentile-based method, it allows for the specification of a pruning ratio, making it more convenient to compare with other baselines. Therefore, to demonstrate the superiority of our method, we have adopted the percentile-based method. More details about the relation between these two methods will be included in the section \ref{pruning_threshold}.

Consequently, the pruned KG can be constructed as: 
\begin{align}
    \Tilde{\mathcal{G}} = \mathcal{G} \odot  \boldsymbol{s} 
\end{align}
where $\boldsymbol{s}$ is the estimated binary mask vector. Finally, our algorithm is summarized in Algorithm \ref{algo_training}.

\begin{algorithm} 
	\caption{Algorithm of KGPR} 
	\label{algo_training} 
        \KwData{users $\mathcal{U}$, items $\mathcal{I}$, user-item interactions $Y$,  and knowledge graph $\mathcal{G}$;}
		\KwIn{batch size $B$, number of layers $L$, balance coefficient $\gamma$, $k$ users per entity;}
        \KwOut{
            pruned knowledge graph $\Tilde{\mathcal{G}}$;
        }
            Construct the user-entity matrix ${Q}$ given the $k$\;
		Initialize learnable embeddings $\boldsymbol{e}_u^{(0)}$, $\boldsymbol{e}_v^{(0)}$, $\boldsymbol{e}_r$ via Xavier Initialization\;
            Assign the learnable parameters $\boldsymbol{e}^h$ to each node and initialize as 1\;
            Set empty list to $\mathcal{S}$ \;
		\While{not convergent}{
		Sample $B$ user-item pairs from $Y$\;
            Construct negative samples $Y^-$ for each interaction and build training data to feed the KGPR model\;
            Lookup the corresponding entities $v$ from $Q$ according to $u$ that appears in the batch\;
            $\boldsymbol{s}^c = sim(\boldsymbol{e}_u^{(0)}, \boldsymbol{e}_v^\prime)$ \;
            Gather the holistic scores $\boldsymbol{s}^h$\;
            $\hat{\boldsymbol{s}}=\gamma \boldsymbol{s}^c + (1-\gamma) \boldsymbol{s}^h$\;
            Append $\hat{\boldsymbol{s}}$ to $\mathcal{S}$\;
            \For{$l \in \left(1,2,\ldots,L\right)$}{
                Calculate the $l$-order representations of $e_v$\;
                $\boldsymbol{e}_v ^{(l)} = KGE(\boldsymbol{e}_v^{(l-1)}, \boldsymbol{e}_r, \hat{\boldsymbol{s}})$\;
                Calculate the $l$-order representation of $e_u$\;
                $\boldsymbol{e}_u^{(l)} =CF (\boldsymbol{e}_u^{(l-1)}, \boldsymbol{e}_i^{(l-1)})$\;
            }
            Calculate the final Embeddings $\boldsymbol{e}_u$ and $\boldsymbol{e}_i$\;
            
            Calculate the BPR loss $\mathcal{L}$\;
		   
		Update $\Theta = \left\{\boldsymbol{e}_u^{(0)}, \boldsymbol{e}_v^{(0)}, \boldsymbol{s}^h\right\}$ according to gradient descent\;
		}
        $\Tilde{\boldsymbol{s}}=AGG(\mathcal{S})$ \;
        Generate the final pruned Graph $\Tilde{\mathcal{G}}$ by percentile-based selection\;
        \Return $\Tilde{\mathcal{G}}$ \;
\end{algorithm}

\subsection{Model Complexity Analysis}
The time cost mainly comes from two parts:  collective importance score calculation and Importance-Aware GNN module.
In the first part, the time expense is greatly reduced because we randomly sampled $k$ users before the calculation. Thus the computational complexity is $O(k|\mathcal{V}|d)$, where $|\mathcal{V}|$ is the number of entities and $d$ is the embedding size.
The second part can further be divided into two phase: collaborative filter phase and KG aggregation phase.
As for collaborative filter phase, the time cost is $O(L|Y|d)$, where $L$ is the number of layer and $|Y|$ is the number of interaction, respectively. Moreover, the computational consumption of KG aggregation phase is $O(L\mathcal{|G|}d)$, where $|\mathcal{G}|$ is the number of triplets. Notably, compared with the cost of the Importance-Aware GNN module, the expense of collective importance score calculation is negligible.


\section{EXPERIMENTS}
\subsection{Experimental Settings}
\subsubsection{Data Description.} To demonstrate the superiority of the proposed KGTrimmer framework, we conduct experiments on three publicly available benchmarking datasets for KGRS, namely Last-FM and Alibaba-Fashion, for comprehensive investigation. The statistics of datasets are summarized in Table \ref{tab:statistics}.
\begin{itemize}
    \item \textbf{Amazon-book\footnote{http://jmcauley.ucsd.edu/data/amazon.}:} This is a popular amazon dataset for product recommendation. 
    In addition,  we refer to \cite{DBLP:conf/cikm/ZhaoHDHOW18} and construct the KG data by mapping items into Freebase entities.
    \item \textbf{Last-FM\footnote{https://grouplens.org/datasets/hetrec-2011.}:} This is a widely-adopted music record dataset released by Last.fm platform, where the tracks are regarded as items. Particularly, the data ranges from Jan 2015 to June 2015 is collected for experiments. Similarly, we align the items with the Freebase entities to fetch the KG data.
    \item \textbf{Alibaba-Fashion\footnote{https://github.com/huangtinglin/Knowledge\_Graph\_based\_Intent\_Network.}:} This is a fashion outfit dataset collected from Alibaba which is an online shopping platform. The outfits are recommended to users as items, where each outfit consists of several fashion stuffs(e.g., tops, bottoms, shoes), and these stuffs are assigned different fashion categories(e.g., jeans, T-shirts). And these attributes are constructed as the KG data of outfits.
\end{itemize}
To ensure the data quality, we use the 10-core setting to guarantee that each user or item has at least 10 interactions. Similarly, the infrequent entities (less than 10 triplets) and relations (less than 50 triplets) are filtered out for the KG part of both datasets. It is worth noting that we utilize the same processed datasets released by KGIN \cite{DBLP:conf/www/WangHWYL0C21} and KGAT \cite{DBLP:conf/kdd/Wang00LC19}, respectively. 

\begin{table}[t]
\caption{Statistics of three real-world datasets.}
\label{tab:statistics}
\setlength{\tabcolsep}{0.4mm}{
\begin{tabular}{cc|c|c|c}
\hline
\multicolumn{2}{c|}{Statistics}                                      & Amazon-book & Last-FM & Alibaba-Fashion  \\
\hline
\multirow{3}{*}{\begin{tabular}[c]{@{}c@{}}User-Item \\ Interactions\end{tabular}} & \# Users      & 70679  & 23,566       & 114,737        \\
                                        & \# Items    & 24915     & 48,123       &   30,040      \\
                                        & \# Interacts & 847733 & 3,034,796       & 1,781,093        \\
\hline
\multirow{3}{*}{\begin{tabular}[c]{@{}c@{}}Knowledge \\ Graph\end{tabular}}        & \# Entities     & 88572 & 58,266       & 59,156        \\
                                        & \# Relations  & 39  & 9       & 51        \\
                                        & \# Triplets    & 2557746 & 464,567       & 279,155        \\
\hline
\end{tabular}
}
\end{table}

\subsubsection{Evaluation Metrics.} In the evaluation phase, we follow the previous studies and adopt the all-ranking strategy for convincing experiments. To measure the performance, two representative evaluation metrics for top-K recommendation, i.e., $Recall@K$ and $NDCG@K$ (Normalized Discounted Cumulative Gain) are adopted for all experiment settings. Note we set $K \in \{10, 20\}$, as small $K$ may encounter the instability of performance, while large $K$ may mismatch the nature of recommendation system that user prefer to clicking the top-K items.
\begin{table*}[t]
\caption{The experimental comparison among a wide range of knowledge graph pruning methods for three benchmarking datasets. Note that \textbf{KGIN} is selected KG-based recommendation model as the backbone. R@K and N@K stand for Recall@K and NDCG@K, respectively. The best results are in bold and the secondary best results are underlined. 
}
\label{tab:overall_performance_kgin}
\setlength{\tabcolsep}{1.5mm}{
\begin{tabular}{p{1.2cm}|c|c|cccccc|c}
\hline

Dataset                              & Ratio                 &  Metric    & Pop & KGRS &CAGED & KGist & MoG & KRDN & KGTrimmer 
\\
\hline
\multirow{12}{1.2cm}{Amazon-book} &  \multirow{4}{*}{90\%} & R@10 & 0.0844    &    \underline{0.1133}     &    0.0736     &  0.0805  & 0.0796& 0.0819   &  \textbf{0.1155}    \\ 
&  & R@20 & 0.1272    &    \underline{0.1641}      &  0.1132     &  0.1196     & 0.1217& 0.1217&  \textbf{0.1667}    \\
&        & N@10 &     0.0544&     \underline{0.0729}     &   0.0467    &  0.0515     & 0.0506  &0.0528&\textbf{0.0741}    \\
&        & N@20 &     0.0675&     \underline{0.0885}       &   0.0587    &  0.0636     & 0.0635&0.0650&  \textbf{0.0898}    \\ 
\cline{2-10}  &         \multirow{4}{*}{95\%} &   R@10   & 0.0860   &  \underline{0.0987}        &  0.0803     &   0.0805    & 0.0786&0.0787&\textbf       {0.1152}     \\
&          &   R@20   & 0.1285   &  \underline{0.1464}        &   0.1204     &   0.1198   &0.1216&0.1177 & \textbf       {0.1662}     \\
&        & N@10 &     0.0553&     \underline{0.0625}        &   0.0504    &  0.0512    &0.0498& 0.0503&  \textbf{0.0733}    \\
&        & N@20 &     0.0683&     \underline{0.0772}        &   0.0627    &  0.0637     &0.0630&0.0622&  \textbf{0.0890}    \\
\cline{2-10}            & \multirow{4}{*}{98\%} &   R@10   &  0.0847  &    \underline{0.0851}      & 0.0833      &   0.0801    &0.0824&0.0805&  \textbf{0.1099}     \\ 
&  &   R@20   &  0.1270  &    \underline{0.1276}        & 0.1260      &   0.1219    &0.1266&0.1210&  \textbf{0.1615}    \\
&         &  N@10    &   \underline{0.0541}  &    0.0540      &  0.0527     &   0.0508    &0.0526&0.0511& \textbf{0.0703} \\
&         &  N@20    &   0.0670  &    \underline{0.0671}      &  0.0659     &   0.0636   &0.0660& 0.0635& \textbf{0.0861}
                         
\\
\hline
\multirow{12}{*}{Last-FM}  & \multirow{4}{*}{30\%} & R@10 & \underline{0.0688}    &    0.0685       &  0.0665     &  0.0656   &0.0647&0.0672&  \textbf{0.0690} \\
 &  & R@20 & 0.0959    &    \underline{0.0963}   &  0.0947     &  0.0925   &0.0922&0.0949&  \textbf{0.0966} \\
&            & N@10 &  \textbf{0.0762}   &   \textbf{0.0762}    &   0.0738    &   0.0728    &0.0725&0.0741&  \underline{0.0758}    \\ 
&            & N@20 &  \underline{0.0838}   &   \textbf{0.0841}  &   0.0818    &   0.0803    &0.0804&0.0819&  \underline{0.0838}    \\ 
\cline{2-10}              & \multirow{4}{*}{70\%} &   R@10   &  0.0603   &    0.0589      &  0.0615     &   0.0597    &0.0560&\underline{0.0612}&   \textbf{0.0667}   \\ 
&  &   R@20   &  0.0856   &    0.0844          &  \underline{0.0885}     &   0.0854    &0.0814&0.0873&   \textbf{0.0936}   \\                         
&      &  N@10    &  0.0675   &  0.0676      &  \underline{0.0692}   &  0.0673     &0.0635&0.0685& \textbf{0.0746} \\
& & N@20    &  0.0743   &  0.0744    &  \underline{0.0768}  &  0.0746     &0.0705&0.0758& \textbf{0.0819} \\
 \cline{2-10}     & \multirow{4}{*}{90\%} &   R@10   &  0.0545   &  \underline{0.0573}      &  0.0562     &   0.0561    &0.0527&0.0553& \textbf{0.0627}     \\
&  &   R@20   &  0.0785   &  0.0818        & \underline{0.0826}     &   0.0804    &0.0766&0.0803& \textbf{0.0889}     \\
&            &  N@10    & 0.0607   &    \underline{0.0649}      &  0.0638     &   0.0630    &0.0603&0.0627& \textbf{0.0720} \\
&            &  N@20    & 0.0675   &    \underline{0.0714}      &  0.0711     &   0.0698    &0.0668&0.0696& \textbf{0.0789}      
\\

\hline

\multirow{12}{1.2cm}{Alibaba-Fashion}  & \multirow{4}{*}{30\%} & R@10 & 0.0761    &    \underline{0.0769}   &  0.0726     &  0.0761   &0.0708&0.0736&  \textbf{0.0773} \\
 &  & R@20 & 0.1161    &    \underline{0.1169}      &  0.1111     &  0.1163   &0.1090&0.1123&  \textbf{0.1179} \\
&            & N@10 &  0.0579   &   \underline{0.0589}     &   0.0551    &   0.0583    &0.0538&0.0562&  \textbf{0.0592}    \\ 
&            & N@20 &  0.0723   &   \underline{0.0732}   &   0.0690    &  0.0728    &0.0675&0.0691&  \textbf{0.0738}    \\ 
\cline{2-10}              & \multirow{4}{*}{70\%} &   R@10   &  \underline{0.0725}   &    0.0690      &   0.0604     &   0.0667    &0.0596&0.0652&   \textbf{0.0761}   \\ 
&  &   R@20   &  \underline{0.1112}   &    0.1070      &    0.0951     &   0.1039    &0.0943&0.1017&   \textbf{0.1154}   \\                         
&      &  N@10    &  \underline{0.0548}   &  0.0515      &  0.0452   &  0.0501     &0.0442&0.0488& \textbf{0.0582} \\
& & N@20    &  \underline{0.0687}   &  0.0652    &  0.0577   &  0.0634     &0.0567&0.0619& \textbf{0.0723} \\
 \cline{2-10}     & \multirow{4}{*}{90\%} &   R@10   &  \underline{0.0656}   &  0.0614        &  0.0592     &   0.0604   &0.0604&0.0623& \textbf{0.0725}     \\
&  &   R@20   &  \underline{0.1018}   &  0.0966        &  0.0930     &   0.0959    &0.0945&0.0970& \textbf{0.1114}     \\
&            &  N@10    & \underline{0.0493}   &    0.0457      &  0.0439     &   0.0446    &0.0445&0.0465& \textbf{0.0553} \\
&            &  N@20    & \underline{0.0623}   &    0.0583     &  0.0560     &   0.0573    &0.0568&0.0589& \textbf{0.0692}  
\\
\hline
\end{tabular}
}

\end{table*}

\begin{table*}[t]
\caption{The experimental comparison among a wide range of knowledge graph pruning methods for three benchmarking datasets. Note that \textbf{KGRec} is selected KG-based recommendation model as the backbone. R@K and N@K stand for Recall@K and NDCG@K, respectively. The best results are in bold and the secondary best results are underlined. 
}
\label{tab:overall_performance_krec}
\setlength{\tabcolsep}{1.5mm}{
\begin{tabular}{p{1.2cm}|c|c|cccccc|c}
\hline

Dataset                              & Ratio                 &  Metric    & Pop & KGRS  & CAGED & KGist & MoG & KRDN & KGTrimmer 
\\
\hline
\multirow{12}{1.2cm}{Amazon-book} &  \multirow{4}{*}{90\%} & R@10 & 0.0861    &    \underline{0.1086}     &    0.0750     &  0.0767 &0.0799&0.0748&  \textbf{0.1111}    \\ 
&  & R@20 & 0.1284    &    \underline{0.1569}     &  0.1138     &  0.1153     &0.1235&0.1136& \textbf{0.1620}    \\
&        & N@10 &     0.0553&     \underline{0.0695}     &    0.0473    &  0.0489     &0.0504&0.0471& \textbf{0.0710}    \\
&        & N@20 &     0.0682&     \underline{0.0845}     &    0.0592    &  0.0607     &0.0636&0.0589&  \textbf{0.0867}    \\ 
\cline{2-10}  &         \multirow{4}{*}{95\%} &   R@10   & 0.0836   &  \underline{0.0947}       &  0.0793     &   0.0776  &0.0772 &0.0700 &\textbf{0.1110}     \\
&          &   R@20   & 0.1251   &  \underline{0.1421}          &  0.1193     &   0.1172   &0.1195&0.1067& \textbf       {0.1625}     \\
&        & N@10 &     0.0535&     \underline{0.0598}        &   0.0500    &  0.0498   &0.0485&0.0441&  \textbf{0.0701}    \\
&        & N@20 &     0.0662&    \underline{0.0743}     &    0.0623    &  0.0619     & 0.0615& 0.0554& \textbf{0.0860}    \\
\cline{2-10}            & \multirow{4}{*}{98\%} &   R@10   &  \underline{0.0865}  &    0.0847         & 0.0840      &   0.0785    & 0.0774& 0.0730&  \textbf{0.1053}     \\ 
&  &   R@20   &  \underline{0.1293}  &    0.1265      &  0.1265      &   0.1185    &0.1199 & 0.1118&  \textbf{0.1557}    \\
&         &  N@10    &   \underline{0.0550}  &    0.0535     &  0.0528     &   0.0498   & 0.0482 & 0.0459 & \textbf{0.0671} \\
&         &  N@20    &   \underline{0.0680}  &    0.0663     &  0.0658     &   0.0621    & 0.0613 & 0.0578 &\textbf{0.0826}
                         
\\
\hline
\multirow{12}{*}{Last-FM}  & \multirow{4}{*}{30\%} & R@10 & \underline{0.0655}    &    \textbf{0.0659}     &  0.0639     &  0.0640   &0.0625&0.0636&  0.0653 \\
 &  & R@20 & \underline{0.0938}    &    \textbf{0.0942}     &  0.0919     &  0.0912   &0.0889&0.0914&  0.0927 \\
&            & N@10 &  \textbf{0.0732}   &   \underline{0.0728}     &   0.0712    &   0.0708    &0.0703&0.0710&  0.0727    \\ 
&            & N@20 &  \textbf{0.0812}   &   \underline{0.0810}   &   0.0789    &   0.0786    &0.0776&0.0789&  0.0803    \\ 
\cline{2-10}              & \multirow{4}{*}{70\%} &   R@10   &  0.0581   &    0.0565    &  0.0579     &   0.0560    &0.0552&\underline{0.0600}&   \textbf{0.0638}   \\ 
&  &   R@20   &  0.0836   &    0.0820        &  0.0844     &   0.0812    &0.0806&\underline{0.0851}&   \textbf{0.0915}   \\                         
&      &  N@10    &  0.0643  &  0.0646       &  0.0654   &  0.0638    &0.0621&\underline{0.0670}& \textbf{0.0717} \\
& & N@20    &  0.0715   &  0.0714    &  0.0727   &  0.0708     &0.0693&\underline{0.0740}& \textbf{0.0793} \\
 \cline{2-10}     & \multirow{4}{*}{90\%} &   R@10   &  0.0535   &  0.0548  & \underline{0.0550}     &   0.0539  &0.0507&0.0540& \textbf{0.0605}     \\
&  &   R@20   &  0.0784   &  0.0786     &  \underline{0.0815}     &   0.0784    &0.0745&0.0793& \textbf{0.0875}     \\
&            &  N@10    & 0.0582   &    \underline{0.0621}     &  0.0615     &   0.0606    &0.0571&0.0608& \textbf{0.0692} \\
&            &  N@20    & 0.0658   &    0.0683   &  \underline{0.0691}    &   0.0675    &0.0638&0.0679& \textbf{0.0764}      
\\

\hline

\multirow{12}{1.2cm}{Alibaba-Fashion}  & \multirow{4}{*}{30\%} & R@10 & \underline{0.0787}    &    \underline{0.0787}   &  0.0747     &  0.0781   &0.0736&0.0758&  \textbf{0.0794} \\
 &  & R@20 & 0.1195    &    \underline{0.1196}    &  0.1135     &  0.1180   &0.1125&0.1157&  \textbf{0.1198} \\
&            & N@10 &  \underline{0.0603}   &   \underline{0.0603}  &   0.0566    &   0.0595    &0.0560&0.0579&  \textbf{0.0609}    \\ 
&            & N@20 &  \underline{0.0749}   &   \underline{0.0749}   &   0.0706    &   0.0739    &0.0700&0.0722& \textbf{0.0754}    \\ 
\cline{2-10}              & \multirow{4}{*}{70\%} &   R@10   &  \underline{0.0731}   &    0.0688  &  0.0636     &   0.0695    &0.0624&0.0668&   \textbf{0.0767}   \\ 
&  &   R@20   &  \underline{0.1117}   &    0.1070      &  0.0993     &   0.1069    &0.0978&0.1039&   \textbf{0.1165}   \\                         
&      &  N@10    &  \underline{0.0558}   &  0.0517      &  0.0478   &  0.0525     &0.0468&0.0503& \textbf{0.0585} \\
& & N@20    & \underline{0.0697}   &  0.0654    &  0.0606   &  0.0659     &0.0595&0.0637& \textbf{0.0729} \\
 \cline{2-10}     & \multirow{4}{*}{90\%} &   R@10   &  \underline{0.0667}   &  0.0634      &  0.0576     &   0.0625    &0.0607&0.0624& \textbf{0.0728}     \\
&  &   R@20   &  \underline{0.1033}   &  0.0991        & 0.1006 &  0.0981    &0.0956&0.0983& \textbf{0.1114}     \\
&            &  N@10    & \underline{0.0500}   &    0.0474     &  0.0424     &  0.0460    &0.0448&0.0463& \textbf{0.0556} \\
&            &  N@20    & \underline{0.0632}   &    0.0603     &  0.0546     &   0.0588    &0.0573&0.0592& \textbf{0.0690}  
\\
\hline
\end{tabular}
}

\end{table*}
                          
                          


\subsubsection{Backbones.} 
To demonstrate the effectiveness and efficiency of the pruned knowledge graph, we employ the following widely-adopted knowledge graph-based recommendation models as the backbone.
\begin{itemize}
    \item \textbf{KGIN} \cite{DBLP:conf/www/WangHWYL0C21}: This work considers user-item interactions at the finer granularity of intents that are a combination of KG relations for better model capacity and interpretability for the recommendation. To this end, the proposed KGIN utilizes a relational path-aware aggregation module to identify the influences of different intents. 
    \item \textbf{KGRec} \cite{DBLP:conf/kdd/YangHXH23}: It is a generative and contrastive self-supervised method to enhance recommender systems by KG. Specifically, KGRec proposes a rational-aware masking mechanism to identify the most important information in the KG, which suppresses potential noise and irrelevant knowledge graph connections.
\end{itemize}
\subsubsection{Compared baselines.} 
Moreover, several pruning methods are also incorporated as baselines to be compared with our KGTrimmer method for recommendation. 
\begin{itemize}
    \item \textbf{Pop:} We sort the entities according to the popularity, and we remove the entities with higher popularity. Here, the in-degree of entities is set as the measure of popularity.
    \item \textbf{KGRS-IMP:} We adopt the GNN framework in this work, and we set $\boldsymbol{s} = \mathbf{1}$ to neglect the importance scores. As a result, we prune the edges based on lower trained embedding norm values.
    \item \textbf{KGist\cite{DBLP:conf/www/BelthZVK20}:} The promising KG summarization method innovatively adopts the Minimum Description Length principle (MDL) to inductively compress the original KG.
    \item \textbf{CAGED\cite{DBLP:conf/cikm/ZhangDDHLX22}:} It is a popular error detection method in KG that predicts the scores of triplets by multi-view contrastive learning. And we finally remove the anomaly entities with connected edges to get the pruned KG.
    \item \textbf{MoG\cite{zhang2024graph}:} This work exerts large effort on graph pruning task, which utilizes multiple sparsifier experts with unique sparsity level and pruning criteria for each node.
    \item \textbf{KRDN\cite{DBLP:conf/sigir/Zhu0M0HG23}:} It is the state-of-the-art denoising knowledge-aware recommendation scheme that tries to distill the high-quality KG triplets and prune the noisy edges. 
\end{itemize}
\subsubsection{Reproductive Settings.}

We implement our KGTrimmer in PyTorch. For a fair comparison, we utilize the Xavier initializer to initialize the embeddings, where the embedding size is fixed to 64 for all methods. We optimize the model parameters with Adam optimizer, where the batch size is set to 4096. The grid search strategy is applied to search hyper-parameters for reaching the optimal performance. Learning rate, regularization coefficient and dropout rate adhere to the settings of previous works\cite{DBLP:conf/www/WangHWYL0C21}. In detail, the learning rate is searched amongst $\{1e^{-5}, 1e^{-4}, 1e^{-3}\}$, the regularization coefficient is tuned in $\{1e^{-6}, 1e^{-5}, 1e^{-4}, 1e^{-3}\}$. As for the GNN structure, the depth of GNN layers is selected from $\{1,2,3\}$, We leverage dropout operations to avoid over-fitting, where the node dropout and message dropout ratios are both tuned amongst $\{0.1, 0.2, ..., 0.7\}$ separately. Furthermore, we choose at most $k\in\{20, 50, 100, 200\}$ users for each entity to construct user-entity matrix.  As for the transform function $\mathcal{T}$, we select from the identity function, average function and multiplicative function. Moreover, we adopt the early stopping strategy by performing premature stopping if the performance does not increase for 100 successive epochs.

\subsection{Experimental Results}
\subsubsection{Overall Performance} In order to demonstrate the superiority of our proposed KGTrimmer methods for knowledge graph recommendation, we compare it with other prune baselines under two state-of-the-art knowledge graph recommendation backbones, and the results are shown in Table \ref{tab:overall_performance_kgin} and \ref{tab:overall_performance_krec}. From the results, we can get several observations. 
\begin{itemize}
    \item  The performance of our proposed KGTrimmer surpasses most of baselines with the same pruning ratio under all datasets, demonstrating the effectiveness in preserving the crucial information of the knowledge graph for recommendation task. It is because collaborative filtering signals enable our model to better capture the influence of entities on users, while other baselines do not consider the role of users in the pruning task. And thanks to the dual-view module, our model can retain the filtered collective information and beneficial holistic information.
    \item The advantages of our method become more evident as the pruning ratio increases. At a pruning ratio of first-class (90\% in Amazon-book, 30\% in Last-FM and Alibaba-Fashion), the performance of a few baselines remains comparable with our method, suggesting that they indeed filter some valueless or noisy entities. However, when pruning more triplets, our method significantly outperforms them because pruning solely relies on the KG itself has reached a performance bottleneck, while the advantages of our model in capturing collaborative filtering signals gradually become prominent.
    \item The larger the Knowledge Graph, the more potential valueless nodes it contains. According to Table \ref{tab:statistics}, the Amazon-book dataset includes 2.5 million triples, substantially more than Last-FM and Alibaba-Fashion, which contain merely 460k and 270k triples, respectively. Interestingly, when the dataset is pruned by 90\%, backbone performance on the Amazon-book remains relatively unaffected, contrasting with the marked performance degradation observed in the other two datasets. This indicates a prevalence of valueless nodes within the Amazon dataset, while the more compact datasets of Last-FM and Alibaba-Fashion appear to contain a higher proportion of valuable information.
    \item Different backbones exhibit varying performance towards the quality of the KG. Specifically, KGIN can achieve better performance compared to KGRec at the same pruning ratio. We attribute this gap to the mask mechanism in KGRec which heightens the reliance on the quality of KG.
    \item CAGED, KGist and MoG are found to be suboptimal in this task, especially in Amazon-book dataset. Although they perform well in their corresponding tasks, they are not originally designed for identifying valueless node and thus fail to recognize the redundant information.
    \item Although KRDN is also a noise-detecting KG-aware recommendation algorithm, its performance in our task was relatively poor. This is because it focuses on obtaining noise-robust representations rather than carefully considering the graph structure. Therefore, it is not suitable for our pruning task.
\end{itemize}


\begin{table}[htbp]
\centering
\caption{Time cost and performance comparison between the original KG and the pruned KG} 
\label{time comparison}
\setlength{\tabcolsep}{1.2mm}{
\begin{tabular}{c|c|c|c|c|c}

\hline
Dataset & Ratio  & Epoch & Total Time & R@20 & N@20 \\ 
\hline
\multirow{2}{*}{Amazon-book}
    & 0\% &  579 & 5.5 days & 0.1688 & 0.0915 \\
    & 90\% &  499 & 0.45 days  & 0.1667 & 0.0898  \\   
\hline
\multirow{2}{*}{Last-FM}
    & 0\%   & 509 & 3.15 days & 0.0978 & 0.0848 \\
    & 30\%   & 489  & 2.31 days & 0.0966 & 0.0838  \\   
\hline
\multirow{2}{*}{Alibaba-Fashion}
    & 0\%  & 209  & 1.06 days & 0.1146 & 0.0716 \\
    & 30\% & 219 & 0.85 days & 0.1179 & 0.0738 \\
\hline

\end{tabular}
}
\end{table}

\subsubsection{Comparison Between Entire KG and the Pruned KG}
The ultimate goal of pruning is to maintain the prediction accuracy while speeding up our prediction, especially in large-scale knowledge graph recommendation tasks. As a result, we compare the pruned knowledge graph generated from KGTrimmer with original one under the KGIN backbone. From the Table \ref{time comparison}, we observe that 
\begin{itemize}
    \item The performance trained on the pruned KG remains comparable with the whole KG, or even surpasses that on Alibaba-Fashion dataset. This observation verifies our motivation that the presence of numerous valueless nodes and noisy nodes in the KG may harm the recommendation performance.
    \item KG pruning can lead to significant time efficiency improvements. It is worth noting that the training efficiency has increased by 12.2 times without any compromise in performance in Amazon-book dataset. Even though the number of user-item interactions is several times higher than the number of triplets in the Last-FM and Alibaba-Fashion dataset, KG pruning still offers substantial time efficiency gains. This phenomenon can be attributed to sophisticated structures designed on the KG in the existing models, which can boost performance but also increase the training time.
    \item Different datasets contain varying number of valueless nodes and noisy nodes. To be specific, in the Amazon-book dataset, performance deterioration only occurs after pruning up to 95\%, while in Last-FM, performance decline is seen at 30\%, and in the Alibaba-Fashion dataset, a relatively good performance is still maintained even after 70\% pruning.
\end{itemize}

\begin{table}[htbp]
\centering
\caption{Comparison among various models on time cost of pruning process} 
\label{tb:cost_pruning_process}
\setlength{\tabcolsep}{1.2mm}{
\begin{tabular}{c|c|c|c}

\hline
 Model & Amazon-book & Last-Fm  & Alibaba-fashion\\ 
\hline
    \multicolumn{4}{l}{Statistical} \\
    \hline
    pop & 12.24s & 5.12s&  3.73s \\
    KGist & 1.08min & 5.56min & 31.10min\\
    \hline 
    \multicolumn{4}{l}{Learning-based} \\
    \hline
    MoG & 5.98h & 50.72min & 35.42min \\
    CAGED &  4.48h & 54.12min   & 30.36h\\
    \hline
    \multicolumn{4}{l}{GNN-based} \\
    \hline
     KRDN & 2.06days & 1.71days & 1.72days\\
    KGRS &  2.00days & 13.27h & 7.41h  \\
     KGTrimmer & 1.43days & 15.62h & 12.05h\\
\hline
\end{tabular}
}
\end{table}
\subsubsection{Time Cost of Pruning Process}
Besides confirming the efficiency of the pruned graph, it is also essential to detail the time complexity of the pruning process. To facilitate clearer distinction among them, we have organized all baselines into three classifications: statistical, learning-based, and GNN-based types. In Table \ref{tb:cost_pruning_process}, we list time cost of baselines and draw the following observations:
\begin{itemize}
    \item The pruning process of statistical methods is very efficient, especially the \textbf{pop} method, which can yield pruning results within one minute. Furthermore, as can be seen in Table \ref{tab:overall_performance_kgin}, the pop method performs well on the Alibaba-Fashion dataset. However, its performance drops significantly on other datasets. Upon deeper investigation, we discovered that the graph structure of the Alibaba-fashion is relatively simple, it primarily consists of outfits and their associated fashion stuffs, such as tops, bottoms, shoes, etc. However, the performance is not as good on more complex graph structures like Last-FM and Amazon-book.
    \item Learning-based methods are notably efficient, typically delivering pruning outcomes within half a day. These methods aim to capture the intricate structures of graphs, resulting in impressive performances across a variety of datasets. Nonetheless, their limitation in identifying the significance of collaborative signals leads to a significant performance gap when compared to our pruning approach.
    \item GNN-based methods are the most time-consuming, generally taking several days to produce results. This can be attributed to the fact that GNNs require stacking multiple layers to capture higher-order signals, a process that takes more time compared to other methods. However, this is also the reason why GNNs usually have stronger expressive capabilities and superior performance. Compared to other GNN-based methods, our KGTrimmer has roughly the same consumption of processing time, but we particularly emphasize the importance of collaborative signals for the pruning task, thus we can produce a higher-quality pruned graph. Additionally, in the future, we will attempt to replace the importance-aware module with a more efficient framework, thus allowing us to obtain the pruned graph within one day or even one hour.
\end{itemize}

\subsubsection{Ablation Study}
In this part, we conduct the ablation study to show how the proposed dual-view importance evaluator affects the final performance of KGTrimmer. To validate the effectiveness of our dual-view Trimmer, we constructed two model variants. In the first variant, we removed the collective score, denoted as \textit{w/o col}. The second variant involves removing the holistic view score, denoted as \textit{w/o hol}. Notably, we conduct experiments on three datasets, employing KGIN as the backbone with a pruning ratio of 98\%, 90\%, 90\% in Amazon-book, Last-Fm, Alibaba-Fashion, repectively, and selecting recall@20 and ndcg@20 as the evaluation metrics.

Based on Table \ref{tab:ablation}, we have made the following discoveries: (1) The model equipped with the dual-view outperformed the other two variants across all datasets, which validates the effectiveness of our overall framework. (2) removing the collective view results in the dramatic decline of overall performance, highlighting the intrinsic importance of the collective view in discerning valueless nodes, which, in turn, substantiates the efficacy of our underlying motivation.
(3) the variant without holistic view also exhibits inferior performance , particularly in Amazon-book dataset, which highlights the importance of capturing the impact of entities on recommendation tasks.

\begin{table}[t]
\caption{Ablation studies that demonstrate the effectiveness of the model designs in Dual-View Importance Evaluator.}
\label{tab:ablation}
\setlength{\tabcolsep}{1mm}{
\begin{tabular}{c|cc|cc|cc}
\hline
                                     & \multicolumn{2}{c|}{Amazon-book}        & \multicolumn{2}{c|}{Last-FM}        & \multicolumn{2}{c}{Alibaba-Fashion}     \\
                                         & Recall & NDCG & Recall & NDCG & Recall & NDCG \\ \hline
 w/o col &  0.1501      &  0.0790    &   0.0824     &   0.0714   &  0.1013      & 0.0611     \\ 
                          w/o hol     &  0.1466      &  0.0778    &  0.0888      &  0.0788    &  0.1083      &  0.0668    \\
                          all            &   \textbf{0.1615}     & \textbf{0.0861}     &  \textbf{0.0889}      &  \textbf{0.0789}    &  \textbf{0.1112}      &  \textbf{0.0694}    
\\ \hline
\end{tabular}
}

\end{table}
\subsubsection{Effect of Pruning Threshold}
\label{pruning_threshold}
During experimental phase, we employ pruning ratio to ensure a fair comparison among baselines. In reality, it is rational to utilize threshold $\tau$ to prune the KG, namely all triplets below $\tau$ will be removed. More detailed discussion could be found in Section \ref{sec:generation}.

\begin{figure}[htbp]
\centering
\caption{The final importance scores' distribution in three datasets. The blue line and red line represent the distribution of entity scores and triplet scores, respectively.}
\label{fig.params.tau}
\vspace{2mm}
\includegraphics[width=1.0\columnwidth]{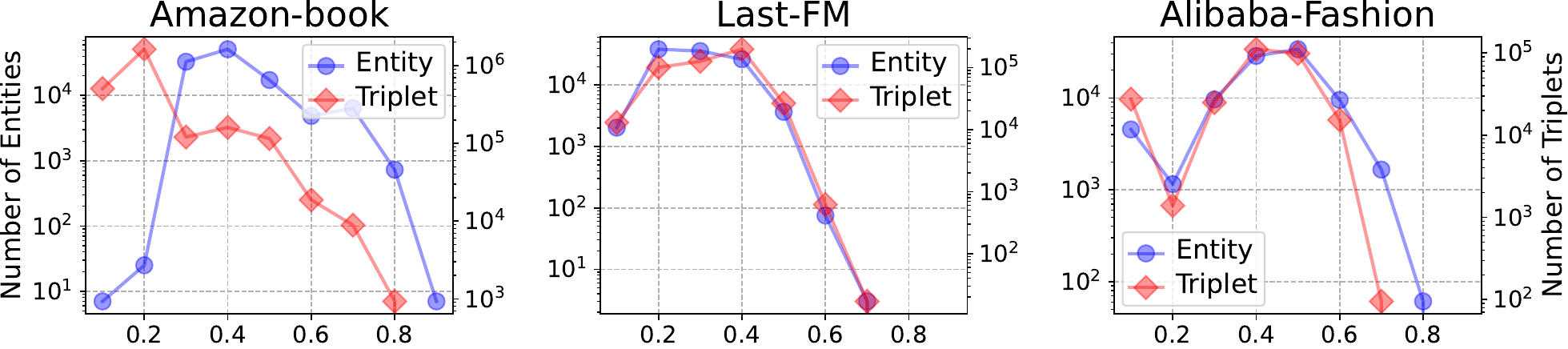}

\end{figure}
To more intuitively display the selection of $\tau$, we have visualized the distribution graph of the final importance score. It should be highlighted that triplet scores are derived from the aggregation of entity scores referring to Equation \ref{score_agg}, such as $\hat{s}(h,r,t)=\hat{s_t}$ , thereby implying that the observed discrepancies in their respective distributions are associated with the degree distribution of the entities. From Figure \ref{fig.params.tau}, we can draw the following observations: (1) All datasets contain vast number of triplets with low scores (less than 0.3), suggesting that non-informative information is a common issue across these datasets. (2) Despite the Amazon-book dataset presenting a limited number of entities with low scores, an unexpectedly large volume of triples have low scores. Examination of the graph structure revealed that seven entities with extremely low score contribute to half of all the triplets in the dataset. This finding confirms that a minority of valueless nodes consume the majority of computational resources.

In summary, selecting an appropriate $\tau$ can be directly applied to these three datasets. If one wish to retain as much useful information as possible, it is feasible to adopt a low $\tau$. Conversely, if one desire faster training efficiency, a larger $\tau$ is a better choice.

\begin{figure}[htbp]
\centering
\caption{Parameters sensitivity to show the how the $\gamma$ affect the performance.}

\label{fig.params.gamma}
 \includegraphics[width=0.95\columnwidth]{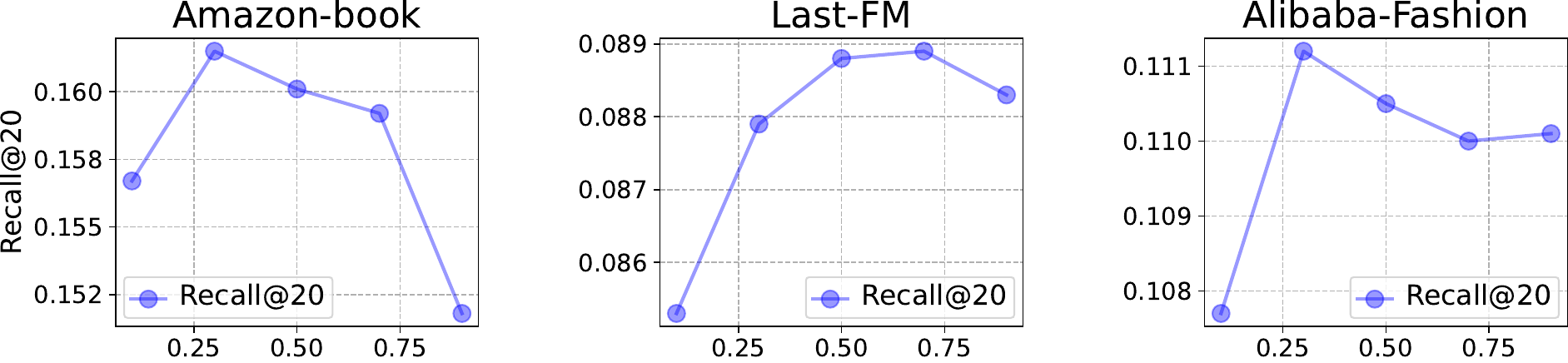}

\vspace{2mm}

\includegraphics[width=0.95\columnwidth]{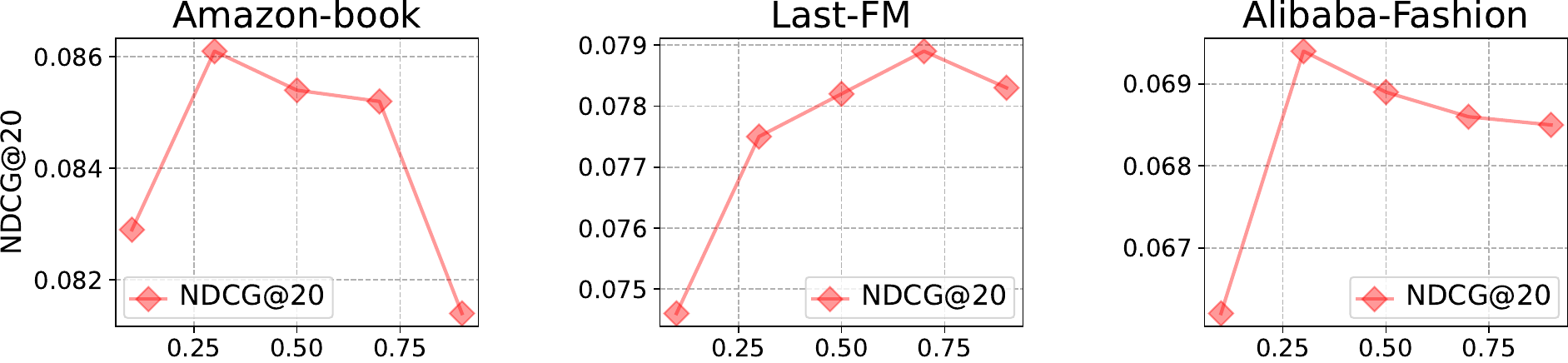}

\label{paramters}
\end{figure}

\subsubsection{Parameters Sensitivity} Moreover, we investigate the sensitivity of hyperparameters in this section. 

\textbf{The impact of $\gamma$.} The parameter $\gamma$ is primarily used to balance the weights between the two views. Specifically, when $\gamma=1$, only the impact of the collective view is considered, whereas when $\gamma=0$ , only the impact of the holistic view is taken into account.
The results of these two aspects have already been demonstrated in the ablation studies. From Figure  \ref{fig.params.gamma}, we have arrived at the following conclusion : (1) Setting $\gamma=0.5$ services as a favorable trade-off for different datasets, indicating that both collective view and holistic view are beneficial to the task. 2) The Amazon-book dataset experiences a dramatically decline at $\gamma$ values of 0.1 and 0.9, indicating that it relies heavily on both the collective and holistic views. Meanwhile, Alibaba-Fashion and Last-FM see a sharp drop at a $\gamma$ value of 0.1, demonstrating that the collective view is of crucial importance for them.

\section{Future Work}
With the exponential growth of data, Large-scale knowledge graphs emerges for its well-organized structure and human-friendly interpretability. However, they lack effective pruning methods which are paving the way for the widespread application of knowledge graph in various fields, such as recommendation. This paper explores the existence of valueless information within knowledge graphs and makes a first attempt at recognizing these nodes via recommendation signals. 

To further approach the ultimate goal, there still exists three challenges need to be addressed: 1) Path-based importance evaluator. In our proposed framework, we focus on acquiring the importance of nodes, while neglecting the potential impact of paths. In fact, we are supposed to comprehensively consider the role played by various information within the paths users take to reach entities to better assess the level of attention these entities receive from users. 2) Efficient importance-aware predictor. Although the GNN-based method proposed in this paper has achieved stunning performance, this module requires substantial computational resources, which hinders the further application in industry filed. Thus, it is urgently needed for efficient implementation of importance-aware predictor. 3) Temporal knowledge graph pruning. Worldwide knowledge is constantly changing, especially in the field of recommendation, where user interests can shift significantly over time. To accommodate swiftly evolving interests, it is essential for the pruning approach to be equipped with the ability to process temporal information. Therefore, effectively integrating temporal knowledge stands as a critical challenge to address in the near future.
\section{Conclusion}
In this paper, we proposed a novel knowledge graph pruning model for recommendation which accelerates the training process and keep the performance stable. Specifically, we first designed a dual-view importance evaluator module to generate aggregated importance scores from collective and holistic views to assess the importance of each entity. Then, based on the importance scores, we constructed the importance-aware graph neural network to build the connection between recommendation system and evaluator module. Then, a new pruned knowledge graph was generated during the training process for downstream recommendation tasks with its lightweight and stable properties. Finally, extensive experiments on three real-world datasets demonstrated the effectiveness of the KGTrimmer compared with competitive baselines. 

\bibliographystyle{ACM-Reference-Format}
\bibliography{sample-base}


\begin{thebibliography}{53}


\ifx \showCODEN    \undefined \def \showCODEN     #1{\unskip}     \fi
\ifx \showDOI      \undefined \def \showDOI       #1{#1}\fi
\ifx \showISBNx    \undefined \def \showISBNx     #1{\unskip}     \fi
\ifx \showISBNxiii \undefined \def \showISBNxiii  #1{\unskip}     \fi
\ifx \showISSN     \undefined \def \showISSN      #1{\unskip}     \fi
\ifx \showLCCN     \undefined \def \showLCCN      #1{\unskip}     \fi
\ifx \shownote     \undefined \def \shownote      #1{#1}          \fi
\ifx \showarticletitle \undefined \def \showarticletitle #1{#1}   \fi
\ifx \showURL      \undefined \def \showURL       {\relax}        \fi
\providecommand\bibfield[2]{#2}
\providecommand\bibinfo[2]{#2}
\providecommand\natexlab[1]{#1}
\providecommand\showeprint[2][]{arXiv:#2}

\bibitem[Ai et~al\mbox{.}(2018)]%
        {DBLP:journals/algorithms/AiACZ18}
\bibfield{author}{\bibinfo{person}{Qingyao Ai}, \bibinfo{person}{Vahid Azizi}, \bibinfo{person}{Xu Chen}, {and} \bibinfo{person}{Yongfeng Zhang}.} \bibinfo{year}{2018}\natexlab{}.
\newblock \showarticletitle{Learning Heterogeneous Knowledge Base Embeddings for Explainable Recommendation}.
\newblock \bibinfo{journal}{\emph{Algorithms}} \bibinfo{volume}{11}, \bibinfo{number}{9} (\bibinfo{year}{2018}), \bibinfo{pages}{137}.
\newblock
\urldef\tempurl%
\url{https://doi.org/10.3390/a11090137}
\showDOI{\tempurl}


\bibitem[Alth{\"{o}}fer et~al\mbox{.}(1990)]%
        {DBLP:conf/swat/AlthoferDDJ90}
\bibfield{author}{\bibinfo{person}{Ingo Alth{\"{o}}fer}, \bibinfo{person}{Gautam Das}, \bibinfo{person}{David~P. Dobkin}, {and} \bibinfo{person}{Deborah Joseph}.} \bibinfo{year}{1990}\natexlab{}.
\newblock \showarticletitle{Generating Sparse Spanners for Weighted Graphs}. In \bibinfo{booktitle}{\emph{{SWAT} 90, 2nd Scandinavian Workshop on Algorithm Theory, Bergen, Norway, July 11-14, 1990, Proceedings}} \emph{(\bibinfo{series}{Lecture Notes in Computer Science}, Vol.~\bibinfo{volume}{447})}, \bibfield{editor}{\bibinfo{person}{John~R. Gilbert} {and} \bibinfo{person}{Rolf~G. Karlsson}} (Eds.). \bibinfo{publisher}{Springer}, \bibinfo{pages}{26--37}.
\newblock
\urldef\tempurl%
\url{https://doi.org/10.1007/3-540-52846-6\_75}
\showDOI{\tempurl}


\bibitem[Batson et~al\mbox{.}(2013)]%
        {DBLP:journals/cacm/BatsonSST13}
\bibfield{author}{\bibinfo{person}{Joshua~D. Batson}, \bibinfo{person}{Daniel~A. Spielman}, \bibinfo{person}{Nikhil Srivastava}, {and} \bibinfo{person}{Shang{-}Hua Teng}.} \bibinfo{year}{2013}\natexlab{}.
\newblock \showarticletitle{Spectral sparsification of graphs: theory and algorithms}.
\newblock \bibinfo{journal}{\emph{Commun. {ACM}}} \bibinfo{volume}{56}, \bibinfo{number}{8} (\bibinfo{year}{2013}), \bibinfo{pages}{87--94}.
\newblock
\urldef\tempurl%
\url{https://doi.org/10.1145/2492007.2492029}
\showDOI{\tempurl}


\bibitem[Belth et~al\mbox{.}(2020)]%
        {DBLP:conf/www/BelthZVK20}
\bibfield{author}{\bibinfo{person}{Caleb Belth}, \bibinfo{person}{Xinyi Zheng}, \bibinfo{person}{Jilles Vreeken}, {and} \bibinfo{person}{Danai Koutra}.} \bibinfo{year}{2020}\natexlab{}.
\newblock \showarticletitle{What is Normal, What is Strange, and What is Missing in a Knowledge Graph: Unified Characterization via Inductive Summarization}. In \bibinfo{booktitle}{\emph{{WWW} '20: The Web Conference 2020, Taipei, Taiwan, April 20-24, 2020}}, \bibfield{editor}{\bibinfo{person}{Yennun Huang}, \bibinfo{person}{Irwin King}, \bibinfo{person}{Tie{-}Yan Liu}, {and} \bibinfo{person}{Maarten van Steen}} (Eds.). \bibinfo{publisher}{{ACM} / {IW3C2}}, \bibinfo{pages}{1115--1126}.
\newblock
\urldef\tempurl%
\url{https://doi.org/10.1145/3366423.3380189}
\showDOI{\tempurl}


\bibitem[Brody et~al\mbox{.}(2022)]%
        {DBLP:conf/iclr/Brody0Y22}
\bibfield{author}{\bibinfo{person}{Shaked Brody}, \bibinfo{person}{Uri Alon}, {and} \bibinfo{person}{Eran Yahav}.} \bibinfo{year}{2022}\natexlab{}.
\newblock \showarticletitle{How Attentive are Graph Attention Networks?}. In \bibinfo{booktitle}{\emph{The Tenth International Conference on Learning Representations, {ICLR} 2022, Virtual Event, April 25-29, 2022}}. \bibinfo{publisher}{OpenReview.net}.
\newblock
\urldef\tempurl%
\url{https://openreview.net/forum?id=F72ximsx7C1}
\showURL{%
\tempurl}


\bibitem[Cao et~al\mbox{.}(2019)]%
        {DBLP:conf/www/0003W0HC19}
\bibfield{author}{\bibinfo{person}{Yixin Cao}, \bibinfo{person}{Xiang Wang}, \bibinfo{person}{Xiangnan He}, \bibinfo{person}{Zikun Hu}, {and} \bibinfo{person}{Tat{-}Seng Chua}.} \bibinfo{year}{2019}\natexlab{}.
\newblock \showarticletitle{Unifying Knowledge Graph Learning and Recommendation: Towards a Better Understanding of User Preferences}. In \bibinfo{booktitle}{\emph{The World Wide Web Conference, {WWW} 2019, San Francisco, CA, USA, May 13-17, 2019}}, \bibfield{editor}{\bibinfo{person}{Ling Liu}, \bibinfo{person}{Ryen~W. White}, \bibinfo{person}{Amin Mantrach}, \bibinfo{person}{Fabrizio Silvestri}, \bibinfo{person}{Julian~J. McAuley}, \bibinfo{person}{Ricardo Baeza{-}Yates}, {and} \bibinfo{person}{Leila Zia}} (Eds.). \bibinfo{publisher}{{ACM}}, \bibinfo{pages}{151--161}.
\newblock
\urldef\tempurl%
\url{https://doi.org/10.1145/3308558.3313705}
\showDOI{\tempurl}


\bibitem[Catherine and Cohen(2016)]%
        {DBLP:conf/recsys/CatherineC16}
\bibfield{author}{\bibinfo{person}{Rose Catherine} {and} \bibinfo{person}{William~W. Cohen}.} \bibinfo{year}{2016}\natexlab{}.
\newblock \showarticletitle{Personalized Recommendations using Knowledge Graphs: {A} Probabilistic Logic Programming Approach}. In \bibinfo{booktitle}{\emph{Proceedings of the 10th {ACM} Conference on Recommender Systems, Boston, MA, USA, September 15-19, 2016}}, \bibfield{editor}{\bibinfo{person}{Shilad Sen}, \bibinfo{person}{Werner Geyer}, \bibinfo{person}{Jill Freyne}, {and} \bibinfo{person}{Pablo Castells}} (Eds.). \bibinfo{publisher}{{ACM}}, \bibinfo{pages}{325--332}.
\newblock
\urldef\tempurl%
\url{https://doi.org/10.1145/2959100.2959131}
\showDOI{\tempurl}


\bibitem[Chen et~al\mbox{.}(2021)]%
        {DBLP:conf/icml/ChenSCZW21}
\bibfield{author}{\bibinfo{person}{Tianlong Chen}, \bibinfo{person}{Yongduo Sui}, \bibinfo{person}{Xuxi Chen}, \bibinfo{person}{Aston Zhang}, {and} \bibinfo{person}{Zhangyang Wang}.} \bibinfo{year}{2021}\natexlab{}.
\newblock \showarticletitle{A Unified Lottery Ticket Hypothesis for Graph Neural Networks}. In \bibinfo{booktitle}{\emph{Proceedings of the 38th International Conference on Machine Learning, {ICML} 2021, 18-24 July 2021, Virtual Event}} \emph{(\bibinfo{series}{Proceedings of Machine Learning Research}, Vol.~\bibinfo{volume}{139})}, \bibfield{editor}{\bibinfo{person}{Marina Meila} {and} \bibinfo{person}{Tong Zhang}} (Eds.). \bibinfo{publisher}{{PMLR}}, \bibinfo{pages}{1695--1706}.
\newblock
\urldef\tempurl%
\url{http://proceedings.mlr.press/v139/chen21p.html}
\showURL{%
\tempurl}


\bibitem[Doerr and Blenn(2013)]%
        {DBLP:conf/sigcomm/DoerrB13}
\bibfield{author}{\bibinfo{person}{Christian Doerr} {and} \bibinfo{person}{Norbert Blenn}.} \bibinfo{year}{2013}\natexlab{}.
\newblock \showarticletitle{Metric convergence in social network sampling}. In \bibinfo{booktitle}{\emph{Proceedings of the 5th {ACM} workshop on Hot topics in planet-scale measurement, HotPlanet@SIGCOMM 2013, Hong Kong, China, August 12-16, 2013}}, \bibfield{editor}{\bibinfo{person}{Pan Hui}, \bibinfo{person}{Emiliano Miluzzo}, {and} \bibinfo{person}{Hamed Haddadi}} (Eds.). \bibinfo{publisher}{{ACM}}, \bibinfo{pages}{45--50}.
\newblock
\urldef\tempurl%
\url{https://doi.org/10.1145/2491159.2491168}
\showDOI{\tempurl}


\bibitem[Du et~al\mbox{.}(2022)]%
        {DBLP:conf/sigir/0002ZCZG22}
\bibfield{author}{\bibinfo{person}{Yuntao Du}, \bibinfo{person}{Xinjun Zhu}, \bibinfo{person}{Lu Chen}, \bibinfo{person}{Baihua Zheng}, {and} \bibinfo{person}{Yunjun Gao}.} \bibinfo{year}{2022}\natexlab{}.
\newblock \showarticletitle{{HAKG:} Hierarchy-Aware Knowledge Gated Network for Recommendation}. In \bibinfo{booktitle}{\emph{{SIGIR} '22: The 45th International {ACM} {SIGIR} Conference on Research and Development in Information Retrieval, Madrid, Spain, July 11 - 15, 2022}}, \bibfield{editor}{\bibinfo{person}{Enrique Amig{\'{o}}}, \bibinfo{person}{Pablo Castells}, \bibinfo{person}{Julio Gonzalo}, \bibinfo{person}{Ben Carterette}, \bibinfo{person}{J.~Shane Culpepper}, {and} \bibinfo{person}{Gabriella Kazai}} (Eds.). \bibinfo{publisher}{{ACM}}, \bibinfo{pages}{1390--1400}.
\newblock
\urldef\tempurl%
\url{https://doi.org/10.1145/3477495.3531987}
\showDOI{\tempurl}


\bibitem[Faralli et~al\mbox{.}(2018)]%
        {DBLP:conf/ijcai/FaralliFPV18}
\bibfield{author}{\bibinfo{person}{Stefano Faralli}, \bibinfo{person}{Irene Finocchi}, \bibinfo{person}{Simone~Paolo Ponzetto}, {and} \bibinfo{person}{Paola Velardi}.} \bibinfo{year}{2018}\natexlab{}.
\newblock \showarticletitle{Efficient Pruning of Large Knowledge Graphs}. In \bibinfo{booktitle}{\emph{Proceedings of the Twenty-Seventh International Joint Conference on Artificial Intelligence, {IJCAI} 2018, July 13-19, 2018, Stockholm, Sweden}}, \bibfield{editor}{\bibinfo{person}{J{\'{e}}r{\^{o}}me Lang}} (Ed.). \bibinfo{publisher}{ijcai.org}, \bibinfo{pages}{4055--4063}.
\newblock
\urldef\tempurl%
\url{https://doi.org/10.24963/ijcai.2018/564}
\showDOI{\tempurl}


\bibitem[Frankle and Carbin(2019)]%
        {DBLP:conf/iclr/FrankleC19}
\bibfield{author}{\bibinfo{person}{Jonathan Frankle} {and} \bibinfo{person}{Michael Carbin}.} \bibinfo{year}{2019}\natexlab{}.
\newblock \showarticletitle{The Lottery Ticket Hypothesis: Finding Sparse, Trainable Neural Networks}. In \bibinfo{booktitle}{\emph{7th International Conference on Learning Representations, {ICLR} 2019, New Orleans, LA, USA, May 6-9, 2019}}. \bibinfo{publisher}{OpenReview.net}.
\newblock
\urldef\tempurl%
\url{https://openreview.net/forum?id=rJl-b3RcF7}
\showURL{%
\tempurl}


\bibitem[Glorot and Bengio(2010)]%
        {DBLP:journals/jmlr/GlorotB10}
\bibfield{author}{\bibinfo{person}{Xavier Glorot} {and} \bibinfo{person}{Yoshua Bengio}.} \bibinfo{year}{2010}\natexlab{}.
\newblock \showarticletitle{Understanding the difficulty of training deep feedforward neural networks}. In \bibinfo{booktitle}{\emph{Proceedings of the Thirteenth International Conference on Artificial Intelligence and Statistics, {AISTATS} 2010, Chia Laguna Resort, Sardinia, Italy, May 13-15, 2010}} \emph{(\bibinfo{series}{{JMLR} Proceedings}, Vol.~\bibinfo{volume}{9})}, \bibfield{editor}{\bibinfo{person}{Yee~Whye Teh} {and} \bibinfo{person}{D.~Mike Titterington}} (Eds.). \bibinfo{publisher}{JMLR.org}, \bibinfo{pages}{249--256}.
\newblock
\urldef\tempurl%
\url{http://proceedings.mlr.press/v9/glorot10a.html}
\showURL{%
\tempurl}


\bibitem[Goyal et~al\mbox{.}(2020)]%
        {DBLP:journals/kbs/GoyalCC20}
\bibfield{author}{\bibinfo{person}{Palash Goyal}, \bibinfo{person}{Sujit~Rokka Chhetri}, {and} \bibinfo{person}{Arquimedes Canedo}.} \bibinfo{year}{2020}\natexlab{}.
\newblock \showarticletitle{dyngraph2vec: Capturing network dynamics using dynamic graph representation learning}.
\newblock \bibinfo{journal}{\emph{Knowl. Based Syst.}}  \bibinfo{volume}{187} (\bibinfo{year}{2020}).
\newblock
\urldef\tempurl%
\url{https://doi.org/10.1016/j.knosys.2019.06.024}
\showDOI{\tempurl}


\bibitem[Hu et~al\mbox{.}(2018)]%
        {DBLP:conf/kdd/HuSZY18}
\bibfield{author}{\bibinfo{person}{Binbin Hu}, \bibinfo{person}{Chuan Shi}, \bibinfo{person}{Wayne~Xin Zhao}, {and} \bibinfo{person}{Philip~S. Yu}.} \bibinfo{year}{2018}\natexlab{}.
\newblock \showarticletitle{Leveraging Meta-path based Context for Top- {N} Recommendation with {A} Neural Co-Attention Model}. In \bibinfo{booktitle}{\emph{Proceedings of the 24th {ACM} {SIGKDD} International Conference on Knowledge Discovery {\&} Data Mining, {KDD} 2018, London, UK, August 19-23, 2018}}, \bibfield{editor}{\bibinfo{person}{Yike Guo} {and} \bibinfo{person}{Faisal Farooq}} (Eds.). \bibinfo{publisher}{{ACM}}, \bibinfo{pages}{1531--1540}.
\newblock
\urldef\tempurl%
\url{https://doi.org/10.1145/3219819.3219965}
\showDOI{\tempurl}


\bibitem[Huang et~al\mbox{.}(2018)]%
        {DBLP:conf/sigir/HuangZDWC18}
\bibfield{author}{\bibinfo{person}{Jin Huang}, \bibinfo{person}{Wayne~Xin Zhao}, \bibinfo{person}{Hongjian Dou}, \bibinfo{person}{Ji{-}Rong Wen}, {and} \bibinfo{person}{Edward~Y. Chang}.} \bibinfo{year}{2018}\natexlab{}.
\newblock \showarticletitle{Improving Sequential Recommendation with Knowledge-Enhanced Memory Networks}. In \bibinfo{booktitle}{\emph{The 41st International {ACM} {SIGIR} Conference on Research {\&} Development in Information Retrieval, {SIGIR} 2018, Ann Arbor, MI, USA, July 08-12, 2018}}, \bibfield{editor}{\bibinfo{person}{Kevyn Collins{-}Thompson}, \bibinfo{person}{Qiaozhu Mei}, \bibinfo{person}{Brian~D. Davison}, \bibinfo{person}{Yiqun Liu}, {and} \bibinfo{person}{Emine Yilmaz}} (Eds.). \bibinfo{publisher}{{ACM}}, \bibinfo{pages}{505--514}.
\newblock
\urldef\tempurl%
\url{https://doi.org/10.1145/3209978.3210017}
\showDOI{\tempurl}


\bibitem[Jacobs et~al\mbox{.}(1991)]%
        {DBLP:journals/neco/JacobsJNH91}
\bibfield{author}{\bibinfo{person}{Robert~A. Jacobs}, \bibinfo{person}{Michael~I. Jordan}, \bibinfo{person}{Steven~J. Nowlan}, {and} \bibinfo{person}{Geoffrey~E. Hinton}.} \bibinfo{year}{1991}\natexlab{}.
\newblock \showarticletitle{Adaptive Mixtures of Local Experts}.
\newblock \bibinfo{journal}{\emph{Neural Comput.}} \bibinfo{volume}{3}, \bibinfo{number}{1} (\bibinfo{year}{1991}), \bibinfo{pages}{79--87}.
\newblock
\urldef\tempurl%
\url{https://doi.org/10.1162/NECO.1991.3.1.79}
\showDOI{\tempurl}


\bibitem[Jia et~al\mbox{.}(2019)]%
        {DBLP:conf/www/JiaXCWE19}
\bibfield{author}{\bibinfo{person}{Shengbin Jia}, \bibinfo{person}{Yang Xiang}, \bibinfo{person}{Xiaojun Chen}, \bibinfo{person}{Kun Wang}, {and} \bibinfo{person}{Shijia E}.} \bibinfo{year}{2019}\natexlab{}.
\newblock \showarticletitle{Triple Trustworthiness Measurement for Knowledge Graph}. In \bibinfo{booktitle}{\emph{The World Wide Web Conference, {WWW} 2019, San Francisco, CA, USA, May 13-17, 2019}}, \bibfield{editor}{\bibinfo{person}{Ling Liu}, \bibinfo{person}{Ryen~W. White}, \bibinfo{person}{Amin Mantrach}, \bibinfo{person}{Fabrizio Silvestri}, \bibinfo{person}{Julian~J. McAuley}, \bibinfo{person}{Ricardo Baeza{-}Yates}, {and} \bibinfo{person}{Leila Zia}} (Eds.). \bibinfo{publisher}{{ACM}}, \bibinfo{pages}{2865--2871}.
\newblock
\urldef\tempurl%
\url{https://doi.org/10.1145/3308558.3313586}
\showDOI{\tempurl}


\bibitem[Jin et~al\mbox{.}(2020)]%
        {DBLP:conf/kdd/JinQFD00ZS20}
\bibfield{author}{\bibinfo{person}{Jiarui Jin}, \bibinfo{person}{Jiarui Qin}, \bibinfo{person}{Yuchen Fang}, \bibinfo{person}{Kounianhua Du}, \bibinfo{person}{Weinan Zhang}, \bibinfo{person}{Yong Yu}, \bibinfo{person}{Zheng Zhang}, {and} \bibinfo{person}{Alexander~J. Smola}.} \bibinfo{year}{2020}\natexlab{}.
\newblock \showarticletitle{An Efficient Neighborhood-based Interaction Model for Recommendation on Heterogeneous Graph}. In \bibinfo{booktitle}{\emph{{KDD} '20: The 26th {ACM} {SIGKDD} Conference on Knowledge Discovery and Data Mining, Virtual Event, CA, USA, August 23-27, 2020}}, \bibfield{editor}{\bibinfo{person}{Rajesh Gupta}, \bibinfo{person}{Yan Liu}, \bibinfo{person}{Jiliang Tang}, {and} \bibinfo{person}{B.~Aditya Prakash}} (Eds.). \bibinfo{publisher}{{ACM}}, \bibinfo{pages}{75--84}.
\newblock
\urldef\tempurl%
\url{https://doi.org/10.1145/3394486.3403050}
\showDOI{\tempurl}


\bibitem[Joshi and Urbani(2020)]%
        {DBLP:conf/www/JoshiU20}
\bibfield{author}{\bibinfo{person}{Unmesh Joshi} {and} \bibinfo{person}{Jacopo Urbani}.} \bibinfo{year}{2020}\natexlab{}.
\newblock \showarticletitle{Searching for Embeddings in a Haystack: Link Prediction on Knowledge Graphs with Subgraph Pruning}. In \bibinfo{booktitle}{\emph{{WWW} '20: The Web Conference 2020, Taipei, Taiwan, April 20-24, 2020}}, \bibfield{editor}{\bibinfo{person}{Yennun Huang}, \bibinfo{person}{Irwin King}, \bibinfo{person}{Tie{-}Yan Liu}, {and} \bibinfo{person}{Maarten van Steen}} (Eds.). \bibinfo{publisher}{{ACM} / {IW3C2}}, \bibinfo{pages}{2817--2823}.
\newblock
\urldef\tempurl%
\url{https://doi.org/10.1145/3366423.3380043}
\showDOI{\tempurl}


\bibitem[Kang et~al\mbox{.}(2022)]%
        {DBLP:conf/icde/KangLS22}
\bibfield{author}{\bibinfo{person}{Shinhwan Kang}, \bibinfo{person}{Kyuhan Lee}, {and} \bibinfo{person}{Kijung Shin}.} \bibinfo{year}{2022}\natexlab{}.
\newblock \showarticletitle{Personalized Graph Summarization: Formulation, Scalable Algorithms, and Applications}. In \bibinfo{booktitle}{\emph{38th {IEEE} International Conference on Data Engineering, {ICDE} 2022, Kuala Lumpur, Malaysia, May 9-12, 2022}}. \bibinfo{publisher}{{IEEE}}, \bibinfo{pages}{2319--2332}.
\newblock
\urldef\tempurl%
\url{https://doi.org/10.1109/ICDE53745.2022.00219}
\showDOI{\tempurl}


\bibitem[Li et~al\mbox{.}(2022)]%
        {DBLP:journals/ijdsa/LiZTJFZ22}
\bibfield{author}{\bibinfo{person}{Jiayu Li}, \bibinfo{person}{Tianyun Zhang}, \bibinfo{person}{Hao Tian}, \bibinfo{person}{Shengmin Jin}, \bibinfo{person}{Makan Fardad}, {and} \bibinfo{person}{Reza Zafarani}.} \bibinfo{year}{2022}\natexlab{}.
\newblock \showarticletitle{Graph sparsification with graph convolutional networks}.
\newblock \bibinfo{journal}{\emph{Int. J. Data Sci. Anal.}} \bibinfo{volume}{13}, \bibinfo{number}{1} (\bibinfo{year}{2022}), \bibinfo{pages}{33--46}.
\newblock
\urldef\tempurl%
\url{https://doi.org/10.1007/S41060-021-00288-8}
\showDOI{\tempurl}


\bibitem[Liu et~al\mbox{.}(2018)]%
        {DBLP:journals/csur/LiuSDK18}
\bibfield{author}{\bibinfo{person}{Yike Liu}, \bibinfo{person}{Tara Safavi}, \bibinfo{person}{Abhilash Dighe}, {and} \bibinfo{person}{Danai Koutra}.} \bibinfo{year}{2018}\natexlab{}.
\newblock \showarticletitle{Graph Summarization Methods and Applications: {A} Survey}.
\newblock \bibinfo{journal}{\emph{{ACM} Comput. Surv.}} \bibinfo{volume}{51}, \bibinfo{number}{3} (\bibinfo{year}{2018}), \bibinfo{pages}{62:1--62:34}.
\newblock
\urldef\tempurl%
\url{https://doi.org/10.1145/3186727}
\showDOI{\tempurl}


\bibitem[Ma et~al\mbox{.}(2019)]%
        {DBLP:conf/www/MaZCJWLMR19}
\bibfield{author}{\bibinfo{person}{Weizhi Ma}, \bibinfo{person}{Min Zhang}, \bibinfo{person}{Yue Cao}, \bibinfo{person}{Woojeong Jin}, \bibinfo{person}{Chenyang Wang}, \bibinfo{person}{Yiqun Liu}, \bibinfo{person}{Shaoping Ma}, {and} \bibinfo{person}{Xiang Ren}.} \bibinfo{year}{2019}\natexlab{}.
\newblock \showarticletitle{Jointly Learning Explainable Rules for Recommendation with Knowledge Graph}. In \bibinfo{booktitle}{\emph{The World Wide Web Conference, {WWW} 2019, San Francisco, CA, USA, May 13-17, 2019}}, \bibfield{editor}{\bibinfo{person}{Ling Liu}, \bibinfo{person}{Ryen~W. White}, \bibinfo{person}{Amin Mantrach}, \bibinfo{person}{Fabrizio Silvestri}, \bibinfo{person}{Julian~J. McAuley}, \bibinfo{person}{Ricardo Baeza{-}Yates}, {and} \bibinfo{person}{Leila Zia}} (Eds.). \bibinfo{publisher}{{ACM}}, \bibinfo{pages}{1210--1221}.
\newblock
\urldef\tempurl%
\url{https://doi.org/10.1145/3308558.3313607}
\showDOI{\tempurl}


\bibitem[Rendle et~al\mbox{.}(2012)]%
        {DBLP:journals/corr/abs-1205-2618}
\bibfield{author}{\bibinfo{person}{Steffen Rendle}, \bibinfo{person}{Christoph Freudenthaler}, \bibinfo{person}{Zeno Gantner}, {and} \bibinfo{person}{Lars Schmidt{-}Thieme}.} \bibinfo{year}{2012}\natexlab{}.
\newblock \showarticletitle{{BPR:} Bayesian Personalized Ranking from Implicit Feedback}.
\newblock \bibinfo{journal}{\emph{CoRR}}  \bibinfo{volume}{abs/1205.2618} (\bibinfo{year}{2012}).
\newblock
\showeprint[arXiv]{1205.2618}
\urldef\tempurl%
\url{http://arxiv.org/abs/1205.2618}
\showURL{%
\tempurl}


\bibitem[Sacenti et~al\mbox{.}(2022)]%
        {DBLP:journals/jiis/SacentiFW22}
\bibfield{author}{\bibinfo{person}{Juarez A.~P. Sacenti}, \bibinfo{person}{Renato Fileto}, {and} \bibinfo{person}{Roberto Willrich}.} \bibinfo{year}{2022}\natexlab{}.
\newblock \showarticletitle{Knowledge graph summarization impacts on movie recommendations}.
\newblock \bibinfo{journal}{\emph{J. Intell. Inf. Syst.}} \bibinfo{volume}{58}, \bibinfo{number}{1} (\bibinfo{year}{2022}), \bibinfo{pages}{43--66}.
\newblock
\urldef\tempurl%
\url{https://doi.org/10.1007/s10844-021-00650-z}
\showDOI{\tempurl}


\bibitem[Shabani et~al\mbox{.}(2023)]%
        {DBLP:journals/corr/abs-2302-06114}
\bibfield{author}{\bibinfo{person}{Nasrin Shabani}, \bibinfo{person}{Jia Wu}, \bibinfo{person}{Amin Beheshti}, \bibinfo{person}{Jin Foo}, \bibinfo{person}{Ambreen Hanif}, {and} \bibinfo{person}{Maryam Shahabikargar}.} \bibinfo{year}{2023}\natexlab{}.
\newblock \showarticletitle{A Survey on Graph Neural Networks for Graph Summarization}.
\newblock \bibinfo{journal}{\emph{CoRR}}  \bibinfo{volume}{abs/2302.06114} (\bibinfo{year}{2023}).
\newblock
\urldef\tempurl%
\url{https://doi.org/10.48550/arXiv.2302.06114}
\showDOI{\tempurl}
\showeprint[arXiv]{2302.06114}


\bibitem[Sun et~al\mbox{.}(2018)]%
        {DBLP:conf/recsys/Sun00BHX18}
\bibfield{author}{\bibinfo{person}{Zhu Sun}, \bibinfo{person}{Jie Yang}, \bibinfo{person}{Jie Zhang}, \bibinfo{person}{Alessandro Bozzon}, \bibinfo{person}{Long{-}Kai Huang}, {and} \bibinfo{person}{Chi Xu}.} \bibinfo{year}{2018}\natexlab{}.
\newblock \showarticletitle{Recurrent knowledge graph embedding for effective recommendation}. In \bibinfo{booktitle}{\emph{Proceedings of the 12th {ACM} Conference on Recommender Systems, RecSys 2018, Vancouver, BC, Canada, October 2-7, 2018}}, \bibfield{editor}{\bibinfo{person}{Sole Pera}, \bibinfo{person}{Michael~D. Ekstrand}, \bibinfo{person}{Xavier Amatriain}, {and} \bibinfo{person}{John O'Donovan}} (Eds.). \bibinfo{publisher}{{ACM}}, \bibinfo{pages}{297--305}.
\newblock
\urldef\tempurl%
\url{https://doi.org/10.1145/3240323.3240361}
\showDOI{\tempurl}


\bibitem[Teru et~al\mbox{.}(2020)]%
        {DBLP:conf/icml/TeruDH20}
\bibfield{author}{\bibinfo{person}{Komal~K. Teru}, \bibinfo{person}{Etienne~G. Denis}, {and} \bibinfo{person}{William~L. Hamilton}.} \bibinfo{year}{2020}\natexlab{}.
\newblock \showarticletitle{Inductive Relation Prediction by Subgraph Reasoning}. In \bibinfo{booktitle}{\emph{Proceedings of the 37th International Conference on Machine Learning, {ICML} 2020, 13-18 July 2020, Virtual Event}} \emph{(\bibinfo{series}{Proceedings of Machine Learning Research}, Vol.~\bibinfo{volume}{119})}. \bibinfo{publisher}{{PMLR}}, \bibinfo{pages}{9448--9457}.
\newblock
\urldef\tempurl%
\url{http://proceedings.mlr.press/v119/teru20a.html}
\showURL{%
\tempurl}


\bibitem[Tian et~al\mbox{.}(2021)]%
        {DBLP:conf/sigir/TianYRWWWL21}
\bibfield{author}{\bibinfo{person}{Yu Tian}, \bibinfo{person}{Yuhao Yang}, \bibinfo{person}{Xudong Ren}, \bibinfo{person}{Pengfei Wang}, \bibinfo{person}{Fangzhao Wu}, \bibinfo{person}{Qian Wang}, {and} \bibinfo{person}{Chenliang Li}.} \bibinfo{year}{2021}\natexlab{}.
\newblock \showarticletitle{Joint Knowledge Pruning and Recurrent Graph Convolution for News Recommendation}. In \bibinfo{booktitle}{\emph{{SIGIR} '21: The 44th International {ACM} {SIGIR} Conference on Research and Development in Information Retrieval, Virtual Event, Canada, July 11-15, 2021}}, \bibfield{editor}{\bibinfo{person}{Fernando Diaz}, \bibinfo{person}{Chirag Shah}, \bibinfo{person}{Torsten Suel}, \bibinfo{person}{Pablo Castells}, \bibinfo{person}{Rosie Jones}, {and} \bibinfo{person}{Tetsuya Sakai}} (Eds.). \bibinfo{publisher}{{ACM}}, \bibinfo{pages}{51--60}.
\newblock
\urldef\tempurl%
\url{https://doi.org/10.1145/3404835.3462912}
\showDOI{\tempurl}


\bibitem[Wang et~al\mbox{.}(2020)]%
        {DBLP:conf/sigir/WangZMLM20}
\bibfield{author}{\bibinfo{person}{Chenyang Wang}, \bibinfo{person}{Min Zhang}, \bibinfo{person}{Weizhi Ma}, \bibinfo{person}{Yiqun Liu}, {and} \bibinfo{person}{Shaoping Ma}.} \bibinfo{year}{2020}\natexlab{}.
\newblock \showarticletitle{Make It a Chorus: Knowledge- and Time-aware Item Modeling for Sequential Recommendation}. In \bibinfo{booktitle}{\emph{Proceedings of the 43rd International {ACM} {SIGIR} conference on research and development in Information Retrieval, {SIGIR} 2020, Virtual Event, China, July 25-30, 2020}}, \bibfield{editor}{\bibinfo{person}{Jimmy~X. Huang}, \bibinfo{person}{Yi~Chang}, \bibinfo{person}{Xueqi Cheng}, \bibinfo{person}{Jaap Kamps}, \bibinfo{person}{Vanessa Murdock}, \bibinfo{person}{Ji{-}Rong Wen}, {and} \bibinfo{person}{Yiqun Liu}} (Eds.). \bibinfo{publisher}{{ACM}}, \bibinfo{pages}{109--118}.
\newblock
\urldef\tempurl%
\url{https://doi.org/10.1145/3397271.3401131}
\showDOI{\tempurl}


\bibitem[Wang et~al\mbox{.}(2018)]%
        {DBLP:conf/cikm/WangZWZLXG18}
\bibfield{author}{\bibinfo{person}{Hongwei Wang}, \bibinfo{person}{Fuzheng Zhang}, \bibinfo{person}{Jialin Wang}, \bibinfo{person}{Miao Zhao}, \bibinfo{person}{Wenjie Li}, \bibinfo{person}{Xing Xie}, {and} \bibinfo{person}{Minyi Guo}.} \bibinfo{year}{2018}\natexlab{}.
\newblock \showarticletitle{RippleNet: Propagating User Preferences on the Knowledge Graph for Recommender Systems}. In \bibinfo{booktitle}{\emph{Proceedings of the 27th {ACM} International Conference on Information and Knowledge Management, {CIKM} 2018, Torino, Italy, October 22-26, 2018}}, \bibfield{editor}{\bibinfo{person}{Alfredo Cuzzocrea}, \bibinfo{person}{James Allan}, \bibinfo{person}{Norman~W. Paton}, \bibinfo{person}{Divesh Srivastava}, \bibinfo{person}{Rakesh Agrawal}, \bibinfo{person}{Andrei~Z. Broder}, \bibinfo{person}{Mohammed~J. Zaki}, \bibinfo{person}{K.~Sel{\c{c}}uk Candan}, \bibinfo{person}{Alexandros Labrinidis}, \bibinfo{person}{Assaf Schuster}, {and} \bibinfo{person}{Haixun Wang}} (Eds.). \bibinfo{publisher}{{ACM}}, \bibinfo{pages}{417--426}.
\newblock
\urldef\tempurl%
\url{https://doi.org/10.1145/3269206.3271739}
\showDOI{\tempurl}


\bibitem[Wang et~al\mbox{.}(2019c)]%
        {DBLP:conf/kdd/WangZZLZLW19}
\bibfield{author}{\bibinfo{person}{Hongwei Wang}, \bibinfo{person}{Fuzheng Zhang}, \bibinfo{person}{Mengdi Zhang}, \bibinfo{person}{Jure Leskovec}, \bibinfo{person}{Miao Zhao}, \bibinfo{person}{Wenjie Li}, {and} \bibinfo{person}{Zhongyuan Wang}.} \bibinfo{year}{2019}\natexlab{c}.
\newblock \showarticletitle{Knowledge-aware Graph Neural Networks with Label Smoothness Regularization for Recommender Systems}. In \bibinfo{booktitle}{\emph{Proceedings of the 25th {ACM} {SIGKDD} International Conference on Knowledge Discovery {\&} Data Mining, {KDD} 2019, Anchorage, AK, USA, August 4-8, 2019}}, \bibfield{editor}{\bibinfo{person}{Ankur Teredesai}, \bibinfo{person}{Vipin Kumar}, \bibinfo{person}{Ying Li}, \bibinfo{person}{R{\'{o}}mer Rosales}, \bibinfo{person}{Evimaria Terzi}, {and} \bibinfo{person}{George Karypis}} (Eds.). \bibinfo{publisher}{{ACM}}, \bibinfo{pages}{968--977}.
\newblock
\urldef\tempurl%
\url{https://doi.org/10.1145/3292500.3330836}
\showDOI{\tempurl}


\bibitem[Wang et~al\mbox{.}(2021d)]%
        {DBLP:journals/access/WangNCL21}
\bibfield{author}{\bibinfo{person}{Jingbin Wang}, \bibinfo{person}{Kuan Nie}, \bibinfo{person}{Xinyuan Chen}, {and} \bibinfo{person}{Jing Lei}.} \bibinfo{year}{2021}\natexlab{d}.
\newblock \showarticletitle{{SUKE:} Embedding Model for Prediction in Uncertain Knowledge Graph}.
\newblock \bibinfo{journal}{\emph{{IEEE} Access}}  \bibinfo{volume}{9} (\bibinfo{year}{2021}), \bibinfo{pages}{3871--3879}.
\newblock
\urldef\tempurl%
\url{https://doi.org/10.1109/ACCESS.2020.3047086}
\showDOI{\tempurl}


\bibitem[Wang et~al\mbox{.}(2021a)]%
        {DBLP:conf/cikm/WangHCWL21}
\bibfield{author}{\bibinfo{person}{Song Wang}, \bibinfo{person}{Xiao Huang}, \bibinfo{person}{Chen Chen}, \bibinfo{person}{Liang Wu}, {and} \bibinfo{person}{Jundong Li}.} \bibinfo{year}{2021}\natexlab{a}.
\newblock \showarticletitle{{REFORM:} Error-Aware Few-Shot Knowledge Graph Completion}. In \bibinfo{booktitle}{\emph{{CIKM} '21: The 30th {ACM} International Conference on Information and Knowledge Management, Virtual Event, Queensland, Australia, November 1 - 5, 2021}}, \bibfield{editor}{\bibinfo{person}{Gianluca Demartini}, \bibinfo{person}{Guido Zuccon}, \bibinfo{person}{J.~Shane Culpepper}, \bibinfo{person}{Zi~Huang}, {and} \bibinfo{person}{Hanghang Tong}} (Eds.). \bibinfo{publisher}{{ACM}}, \bibinfo{pages}{1979--1988}.
\newblock
\urldef\tempurl%
\url{https://doi.org/10.1145/3459637.3482470}
\showDOI{\tempurl}


\bibitem[Wang et~al\mbox{.}(2019a)]%
        {DBLP:conf/kdd/Wang00LC19}
\bibfield{author}{\bibinfo{person}{Xiang Wang}, \bibinfo{person}{Xiangnan He}, \bibinfo{person}{Yixin Cao}, \bibinfo{person}{Meng Liu}, {and} \bibinfo{person}{Tat{-}Seng Chua}.} \bibinfo{year}{2019}\natexlab{a}.
\newblock \showarticletitle{{KGAT:} Knowledge Graph Attention Network for Recommendation}. In \bibinfo{booktitle}{\emph{Proceedings of the 25th {ACM} {SIGKDD} International Conference on Knowledge Discovery {\&} Data Mining, {KDD} 2019, Anchorage, AK, USA, August 4-8, 2019}}, \bibfield{editor}{\bibinfo{person}{Ankur Teredesai}, \bibinfo{person}{Vipin Kumar}, \bibinfo{person}{Ying Li}, \bibinfo{person}{R{\'{o}}mer Rosales}, \bibinfo{person}{Evimaria Terzi}, {and} \bibinfo{person}{George Karypis}} (Eds.). \bibinfo{publisher}{{ACM}}, \bibinfo{pages}{950--958}.
\newblock
\urldef\tempurl%
\url{https://doi.org/10.1145/3292500.3330989}
\showDOI{\tempurl}


\bibitem[Wang et~al\mbox{.}(2021b)]%
        {DBLP:conf/www/WangHWYL0C21}
\bibfield{author}{\bibinfo{person}{Xiang Wang}, \bibinfo{person}{Tinglin Huang}, \bibinfo{person}{Dingxian Wang}, \bibinfo{person}{Yancheng Yuan}, \bibinfo{person}{Zhenguang Liu}, \bibinfo{person}{Xiangnan He}, {and} \bibinfo{person}{Tat{-}Seng Chua}.} \bibinfo{year}{2021}\natexlab{b}.
\newblock \showarticletitle{Learning Intents behind Interactions with Knowledge Graph for Recommendation}. In \bibinfo{booktitle}{\emph{{WWW} '21: The Web Conference 2021, Virtual Event / Ljubljana, Slovenia, April 19-23, 2021}}, \bibfield{editor}{\bibinfo{person}{Jure Leskovec}, \bibinfo{person}{Marko Grobelnik}, \bibinfo{person}{Marc Najork}, \bibinfo{person}{Jie Tang}, {and} \bibinfo{person}{Leila Zia}} (Eds.). \bibinfo{publisher}{{ACM} / {IW3C2}}, \bibinfo{pages}{878--887}.
\newblock
\urldef\tempurl%
\url{https://doi.org/10.1145/3442381.3450133}
\showDOI{\tempurl}


\bibitem[Wang et~al\mbox{.}(2019b)]%
        {DBLP:conf/aaai/WangWX00C19}
\bibfield{author}{\bibinfo{person}{Xiang Wang}, \bibinfo{person}{Dingxian Wang}, \bibinfo{person}{Canran Xu}, \bibinfo{person}{Xiangnan He}, \bibinfo{person}{Yixin Cao}, {and} \bibinfo{person}{Tat{-}Seng Chua}.} \bibinfo{year}{2019}\natexlab{b}.
\newblock \showarticletitle{Explainable Reasoning over Knowledge Graphs for Recommendation}. In \bibinfo{booktitle}{\emph{The Thirty-Third {AAAI} Conference on Artificial Intelligence, {AAAI} 2019, The Thirty-First Innovative Applications of Artificial Intelligence Conference, {IAAI} 2019, The Ninth {AAAI} Symposium on Educational Advances in Artificial Intelligence, {EAAI} 2019, Honolulu, Hawaii, USA, January 27 - February 1, 2019}}. \bibinfo{publisher}{{AAAI} Press}, \bibinfo{pages}{5329--5336}.
\newblock
\urldef\tempurl%
\url{https://doi.org/10.1609/aaai.v33i01.33015329}
\showDOI{\tempurl}


\bibitem[Wang et~al\mbox{.}(2021c)]%
        {DBLP:conf/cikm/WangLFSY21}
\bibfield{author}{\bibinfo{person}{Yu Wang}, \bibinfo{person}{Zhiwei Liu}, \bibinfo{person}{Ziwei Fan}, \bibinfo{person}{Lichao Sun}, {and} \bibinfo{person}{Philip~S. Yu}.} \bibinfo{year}{2021}\natexlab{c}.
\newblock \showarticletitle{DSKReG: Differentiable Sampling on Knowledge Graph for Recommendation with Relational {GNN}}. In \bibinfo{booktitle}{\emph{{CIKM} '21: The 30th {ACM} International Conference on Information and Knowledge Management, Virtual Event, Queensland, Australia, November 1 - 5, 2021}}, \bibfield{editor}{\bibinfo{person}{Gianluca Demartini}, \bibinfo{person}{Guido Zuccon}, \bibinfo{person}{J.~Shane Culpepper}, \bibinfo{person}{Zi~Huang}, {and} \bibinfo{person}{Hanghang Tong}} (Eds.). \bibinfo{publisher}{{ACM}}, \bibinfo{pages}{3513--3517}.
\newblock
\urldef\tempurl%
\url{https://doi.org/10.1145/3459637.3482092}
\showDOI{\tempurl}


\bibitem[Wang et~al\mbox{.}(2014)]%
        {DBLP:conf/aaai/WangZFC14}
\bibfield{author}{\bibinfo{person}{Zhen Wang}, \bibinfo{person}{Jianwen Zhang}, \bibinfo{person}{Jianlin Feng}, {and} \bibinfo{person}{Zheng Chen}.} \bibinfo{year}{2014}\natexlab{}.
\newblock \showarticletitle{Knowledge Graph Embedding by Translating on Hyperplanes}. In \bibinfo{booktitle}{\emph{Proceedings of the Twenty-Eighth {AAAI} Conference on Artificial Intelligence, July 27 -31, 2014, Qu{\'{e}}bec City, Qu{\'{e}}bec, Canada}}, \bibfield{editor}{\bibinfo{person}{Carla~E. Brodley} {and} \bibinfo{person}{Peter Stone}} (Eds.). \bibinfo{publisher}{{AAAI} Press}, \bibinfo{pages}{1112--1119}.
\newblock
\urldef\tempurl%
\url{https://doi.org/10.1609/AAAI.V28I1.8870}
\showDOI{\tempurl}


\bibitem[Wei et~al\mbox{.}(2022)]%
        {DBLP:journals/corr/abs-2212-10046}
\bibfield{author}{\bibinfo{person}{Yinwei Wei}, \bibinfo{person}{Xiang Wang}, \bibinfo{person}{Liqiang Nie}, \bibinfo{person}{Shaoyu Li}, \bibinfo{person}{Dingxian Wang}, {and} \bibinfo{person}{Tat{-}Seng Chua}.} \bibinfo{year}{2022}\natexlab{}.
\newblock \showarticletitle{Causal Inference for Knowledge Graph based Recommendation}.
\newblock \bibinfo{journal}{\emph{CoRR}}  \bibinfo{volume}{abs/2212.10046} (\bibinfo{year}{2022}).
\newblock
\urldef\tempurl%
\url{https://doi.org/10.48550/arXiv.2212.10046}
\showDOI{\tempurl}
\showeprint[arXiv]{2212.10046}


\bibitem[Xie et~al\mbox{.}(2018)]%
        {DBLP:conf/aaai/Xie0LL18}
\bibfield{author}{\bibinfo{person}{Ruobing Xie}, \bibinfo{person}{Zhiyuan Liu}, \bibinfo{person}{Fen Lin}, {and} \bibinfo{person}{Leyu Lin}.} \bibinfo{year}{2018}\natexlab{}.
\newblock \showarticletitle{Does William Shakespeare {REALLY} Write Hamlet? Knowledge Representation Learning With Confidence}. In \bibinfo{booktitle}{\emph{Proceedings of the Thirty-Second {AAAI} Conference on Artificial Intelligence, (AAAI-18), the 30th innovative Applications of Artificial Intelligence (IAAI-18), and the 8th {AAAI} Symposium on Educational Advances in Artificial Intelligence (EAAI-18), New Orleans, Louisiana, USA, February 2-7, 2018}}, \bibfield{editor}{\bibinfo{person}{Sheila~A. McIlraith} {and} \bibinfo{person}{Kilian~Q. Weinberger}} (Eds.). \bibinfo{publisher}{{AAAI} Press}, \bibinfo{pages}{4954--4961}.
\newblock
\urldef\tempurl%
\url{https://doi.org/10.1609/AAAI.V32I1.11924}
\showDOI{\tempurl}


\bibitem[Xu et~al\mbox{.}(2020)]%
        {DBLP:conf/iclr/XuFJXSD20}
\bibfield{author}{\bibinfo{person}{Xiaoran Xu}, \bibinfo{person}{Wei Feng}, \bibinfo{person}{Yunsheng Jiang}, \bibinfo{person}{Xiaohui Xie}, \bibinfo{person}{Zhiqing Sun}, {and} \bibinfo{person}{Zhi{-}Hong Deng}.} \bibinfo{year}{2020}\natexlab{}.
\newblock \showarticletitle{Dynamically Pruned Message Passing Networks for Large-scale Knowledge Graph Reasoning}. In \bibinfo{booktitle}{\emph{8th International Conference on Learning Representations, {ICLR} 2020, Addis Ababa, Ethiopia, April 26-30, 2020}}. \bibinfo{publisher}{OpenReview.net}.
\newblock
\urldef\tempurl%
\url{https://openreview.net/forum?id=rkeuAhVKvB}
\showURL{%
\tempurl}


\bibitem[Xu et~al\mbox{.}(2022)]%
        {DBLP:conf/www/XuDT22}
\bibfield{author}{\bibinfo{person}{Zhe Xu}, \bibinfo{person}{Boxin Du}, {and} \bibinfo{person}{Hanghang Tong}.} \bibinfo{year}{2022}\natexlab{}.
\newblock \showarticletitle{Graph Sanitation with Application to Node Classification}. In \bibinfo{booktitle}{\emph{{WWW} '22: The {ACM} Web Conference 2022, Virtual Event, Lyon, France, April 25 - 29, 2022}}, \bibfield{editor}{\bibinfo{person}{Fr{\'{e}}d{\'{e}}rique Laforest}, \bibinfo{person}{Rapha{\"{e}}l Troncy}, \bibinfo{person}{Elena Simperl}, \bibinfo{person}{Deepak Agarwal}, \bibinfo{person}{Aristides Gionis}, \bibinfo{person}{Ivan Herman}, {and} \bibinfo{person}{Lionel M{\'{e}}dini}} (Eds.). \bibinfo{publisher}{{ACM}}, \bibinfo{pages}{1136--1147}.
\newblock
\urldef\tempurl%
\url{https://doi.org/10.1145/3485447.3512180}
\showDOI{\tempurl}


\bibitem[Yang et~al\mbox{.}(2023)]%
        {DBLP:conf/kdd/YangHXH23}
\bibfield{author}{\bibinfo{person}{Yuhao Yang}, \bibinfo{person}{Chao Huang}, \bibinfo{person}{Lianghao Xia}, {and} \bibinfo{person}{Chunzhen Huang}.} \bibinfo{year}{2023}\natexlab{}.
\newblock \showarticletitle{Knowledge Graph Self-Supervised Rationalization for Recommendation}. In \bibinfo{booktitle}{\emph{Proceedings of the 29th {ACM} {SIGKDD} Conference on Knowledge Discovery and Data Mining, {KDD} 2023, Long Beach, CA, USA, August 6-10, 2023}}, \bibfield{editor}{\bibinfo{person}{Ambuj~K. Singh}, \bibinfo{person}{Yizhou Sun}, \bibinfo{person}{Leman Akoglu}, \bibinfo{person}{Dimitrios Gunopulos}, \bibinfo{person}{Xifeng Yan}, \bibinfo{person}{Ravi Kumar}, \bibinfo{person}{Fatma Ozcan}, {and} \bibinfo{person}{Jieping Ye}} (Eds.). \bibinfo{publisher}{{ACM}}, \bibinfo{pages}{3046--3056}.
\newblock
\urldef\tempurl%
\url{https://doi.org/10.1145/3580305.3599400}
\showDOI{\tempurl}


\bibitem[Yu et~al\mbox{.}(2022)]%
        {DBLP:conf/uai/YuAYJP22}
\bibfield{author}{\bibinfo{person}{Shujian Yu}, \bibinfo{person}{Francesco Alesiani}, \bibinfo{person}{Wenzhe Yin}, \bibinfo{person}{Robert Jenssen}, {and} \bibinfo{person}{Jos{\'{e}}~C. Pr{\'{\i}}ncipe}.} \bibinfo{year}{2022}\natexlab{}.
\newblock \showarticletitle{Principle of relevant information for graph sparsification}. In \bibinfo{booktitle}{\emph{Uncertainty in Artificial Intelligence, Proceedings of the Thirty-Eighth Conference on Uncertainty in Artificial Intelligence, {UAI} 2022, 1-5 August 2022, Eindhoven, The Netherlands}} \emph{(\bibinfo{series}{Proceedings of Machine Learning Research}, Vol.~\bibinfo{volume}{180})}, \bibfield{editor}{\bibinfo{person}{James Cussens} {and} \bibinfo{person}{Kun Zhang}} (Eds.). \bibinfo{publisher}{{PMLR}}, \bibinfo{pages}{2331--2341}.
\newblock
\urldef\tempurl%
\url{https://proceedings.mlr.press/v180/yu22c.html}
\showURL{%
\tempurl}


\bibitem[Zhang et~al\mbox{.}(2016)]%
        {DBLP:conf/kdd/ZhangYLXM16}
\bibfield{author}{\bibinfo{person}{Fuzheng Zhang}, \bibinfo{person}{Nicholas~Jing Yuan}, \bibinfo{person}{Defu Lian}, \bibinfo{person}{Xing Xie}, {and} \bibinfo{person}{Wei{-}Ying Ma}.} \bibinfo{year}{2016}\natexlab{}.
\newblock \showarticletitle{Collaborative Knowledge Base Embedding for Recommender Systems}. In \bibinfo{booktitle}{\emph{Proceedings of the 22nd {ACM} {SIGKDD} International Conference on Knowledge Discovery and Data Mining, San Francisco, CA, USA, August 13-17, 2016}}, \bibfield{editor}{\bibinfo{person}{Balaji Krishnapuram}, \bibinfo{person}{Mohak Shah}, \bibinfo{person}{Alexander~J. Smola}, \bibinfo{person}{Charu~C. Aggarwal}, \bibinfo{person}{Dou Shen}, {and} \bibinfo{person}{Rajeev Rastogi}} (Eds.). \bibinfo{publisher}{{ACM}}, \bibinfo{pages}{353--362}.
\newblock
\urldef\tempurl%
\url{https://doi.org/10.1145/2939672.2939673}
\showDOI{\tempurl}


\bibitem[Zhang et~al\mbox{.}(2024a)]%
        {zhang2024graph}
\bibfield{author}{\bibinfo{person}{Guibin Zhang}, \bibinfo{person}{Xiangguo Sun}, \bibinfo{person}{Yanwei Yue}, \bibinfo{person}{Kun Wang}, \bibinfo{person}{Tianlong Chen}, {and} \bibinfo{person}{Shirui Pan}.} \bibinfo{year}{2024}\natexlab{a}.
\newblock \showarticletitle{Graph Sparsification via Mixture of Graphs}.
\newblock \bibinfo{journal}{\emph{arXiv preprint arXiv:2405.14260}} (\bibinfo{year}{2024}).
\newblock


\bibitem[Zhang et~al\mbox{.}(2024b)]%
        {zhang2024graph1}
\bibfield{author}{\bibinfo{person}{Guibin Zhang}, \bibinfo{person}{Kun Wang}, \bibinfo{person}{Wei Huang}, \bibinfo{person}{Yanwei Yue}, \bibinfo{person}{Yang Wang}, \bibinfo{person}{Roger Zimmermann}, \bibinfo{person}{Aojun Zhou}, \bibinfo{person}{Dawei Cheng}, \bibinfo{person}{Jin Zeng}, {and} \bibinfo{person}{Yuxuan Liang}.} \bibinfo{year}{2024}\natexlab{b}.
\newblock \showarticletitle{Graph lottery ticket automated}. In \bibinfo{booktitle}{\emph{The Twelfth International Conference on Learning Representations}}.
\newblock


\bibitem[Zhang et~al\mbox{.}(2022)]%
        {DBLP:conf/cikm/ZhangDDHLX22}
\bibfield{author}{\bibinfo{person}{Qinggang Zhang}, \bibinfo{person}{Junnan Dong}, \bibinfo{person}{Keyu Duan}, \bibinfo{person}{Xiao Huang}, \bibinfo{person}{Yezi Liu}, {and} \bibinfo{person}{Linchuan Xu}.} \bibinfo{year}{2022}\natexlab{}.
\newblock \showarticletitle{Contrastive Knowledge Graph Error Detection}. In \bibinfo{booktitle}{\emph{Proceedings of the 31st {ACM} International Conference on Information {\&} Knowledge Management, Atlanta, GA, USA, October 17-21, 2022}}, \bibfield{editor}{\bibinfo{person}{Mohammad~Al Hasan} {and} \bibinfo{person}{Li~Xiong}} (Eds.). \bibinfo{publisher}{{ACM}}, \bibinfo{pages}{2590--2599}.
\newblock
\urldef\tempurl%
\url{https://doi.org/10.1145/3511808.3557264}
\showDOI{\tempurl}


\bibitem[Zhao et~al\mbox{.}(2018)]%
        {DBLP:conf/cikm/ZhaoHDHOW18}
\bibfield{author}{\bibinfo{person}{Wayne~Xin Zhao}, \bibinfo{person}{Gaole He}, \bibinfo{person}{Hongjian Dou}, \bibinfo{person}{Jin Huang}, \bibinfo{person}{Siqi Ouyang}, {and} \bibinfo{person}{Ji{-}Rong Wen}.} \bibinfo{year}{2018}\natexlab{}.
\newblock \showarticletitle{KB4Rec: {A} Dataset for Linking Knowledge Bases with Recommender Systems}. In \bibinfo{booktitle}{\emph{Proceedings of the {CIKM} 2018 Workshops co-located with 27th {ACM} International Conference on Information and Knowledge Management {(CIKM} 2018), Torino, Italy, October 22, 2018}} \emph{(\bibinfo{series}{{CEUR} Workshop Proceedings}, Vol.~\bibinfo{volume}{2482})}, \bibfield{editor}{\bibinfo{person}{Alfredo Cuzzocrea}, \bibinfo{person}{Francesco Bonchi}, {and} \bibinfo{person}{Dimitrios Gunopulos}} (Eds.). \bibinfo{publisher}{CEUR-WS.org}.
\newblock
\urldef\tempurl%
\url{https://ceur-ws.org/Vol-2482/paper25.pdf}
\showURL{%
\tempurl}


\bibitem[Zhou et~al\mbox{.}(2020)]%
        {DBLP:journals/aiopen/ZhouCHZYLWLS20}
\bibfield{author}{\bibinfo{person}{Jie Zhou}, \bibinfo{person}{Ganqu Cui}, \bibinfo{person}{Shengding Hu}, \bibinfo{person}{Zhengyan Zhang}, \bibinfo{person}{Cheng Yang}, \bibinfo{person}{Zhiyuan Liu}, \bibinfo{person}{Lifeng Wang}, \bibinfo{person}{Changcheng Li}, {and} \bibinfo{person}{Maosong Sun}.} \bibinfo{year}{2020}\natexlab{}.
\newblock \showarticletitle{Graph neural networks: {A} review of methods and applications}.
\newblock \bibinfo{journal}{\emph{{AI} Open}}  \bibinfo{volume}{1} (\bibinfo{year}{2020}), \bibinfo{pages}{57--81}.
\newblock
\urldef\tempurl%
\url{https://doi.org/10.1016/j.aiopen.2021.01.001}
\showDOI{\tempurl}


\bibitem[Zhu et~al\mbox{.}(2023)]%
        {DBLP:conf/sigir/Zhu0M0HG23}
\bibfield{author}{\bibinfo{person}{Xinjun Zhu}, \bibinfo{person}{Yuntao Du}, \bibinfo{person}{Yuren Mao}, \bibinfo{person}{Lu Chen}, \bibinfo{person}{Yujia Hu}, {and} \bibinfo{person}{Yunjun Gao}.} \bibinfo{year}{2023}\natexlab{}.
\newblock \showarticletitle{Knowledge-refined Denoising Network for Robust Recommendation}. In \bibinfo{booktitle}{\emph{Proceedings of the 46th International {ACM} {SIGIR} Conference on Research and Development in Information Retrieval, {SIGIR} 2023, Taipei, Taiwan, July 23-27, 2023}}, \bibfield{editor}{\bibinfo{person}{Hsin{-}Hsi Chen}, \bibinfo{person}{Wei{-}Jou~(Edward) Duh}, \bibinfo{person}{Hen{-}Hsen Huang}, \bibinfo{person}{Makoto~P. Kato}, \bibinfo{person}{Josiane Mothe}, {and} \bibinfo{person}{Barbara Poblete}} (Eds.). \bibinfo{publisher}{{ACM}}, \bibinfo{pages}{362--371}.
\newblock
\urldef\tempurl%
\url{https://doi.org/10.1145/3539618.3591707}
\showDOI{\tempurl}


\end{thebibliography}

\appendix

\end{document}